\documentclass[acmlarge]{acmart}

\usepackage[ruled,vlined]{algorithm2e}
\usepackage{amsmath}
\usepackage{amsthm}
\usepackage{amssymb}
\usepackage{color, colortbl}
\usepackage{subfigure}
\newcommand{\nop}[1]{}
\definecolor{MyBlue}{RGB}{10,80,180}
\usepackage[normalem]{ulem}
\usepackage{soul}
\usepackage{geometry}
\usepackage{graphicx}

\setstcolor{red}

\nop{
\usepackage{draftwatermark}
\SetWatermarkText{Under Review. Do Not Distribute.}
\SetWatermarkScale{0.4}
}

\AtBeginDocument{%
  \providecommand\BibTeX{{%
    \normalfont B\kern-0.5em{\scshape i\kern-0.25em b}\kern-0.8em\TeX}}}

\nop{
\setcopyright{acmcopyright}
\copyrightyear{2019}
\acmYear{2019}
\acmDOI{10.1145/1122445.1122456}
}

\acmJournal{IMWUT}
\acmVolume{1}
\acmNumber{1}
\acmArticle{1}
\acmMonth{11}

\begin{document}

\title{Quantifying Low-Battery Anxiety of Mobile Users and Its Impacts on Video Watching Behavior}

\author{Guoming Tang}
\authornote{Corresponding author: Guoming Tang (tanggm@pcl.ac.cn).}
\affiliation{%
  \institution{Peng Cheng Laboratory}
  \city{Shenzhen, Guangdong}
  \country{P. R. China}}

\author{Kui Wu}
\affiliation{%
  \institution{University of Victoria}
  \city{Victoria, BC}
  \country{Canada}}

\author{Yangjing Wu}
\affiliation{%
  \institution{Sandalwood Advisors}
  \city{Hong Kong}
  \country{P. R. China}}

\author{Hanlong Liao}
\affiliation{%
  \institution{National University of Defense Technology}
  \city{Changsha, Hunan}
  \country{P. R. China}}

\author{Deke Guo}
\affiliation{%
  \institution{National University of Defense Technology}
  \city{Changsha, Hunan}
  \country{P. R. China}}

\author{Yi Wang}
\affiliation{%
  \institution{Peng Cheng Laboratory; Southern University of Science and Technology}
  \city{Shenzhen, Guangdong}
  \country{P. R. China}}

\nop{
\renewcommand{\shortauthors}{G. Tang et al.}
}
\begin{abstract}
People nowadays are increasingly dependent on mobile phones for daily communication, study, and business. Along with this it incurs the low-battery anxiety (LBA). Although having been unveiled for a while, LBA has not been thoroughly investigated yet. Without a better understanding of LBA, it would be difficult to precisely validate energy saving and management techniques in terms of alleviating LBA and enhancing Quality of Experience (QoE) of mobile users. To fill the gap, we conduct an investigation over $2000+$ mobile users, look into their feelings and reactions towards LBA, and quantify their anxiety degree during the draining of battery power. As a case study, we also investigate the impact of LBA on user's behavior at video watching, and with the massive collected answers we are able to quantify user's abandoning likelihood of attractive videos versus the battery status of mobile phone. The empirical findings and quantitative models obtained in this work not only disclose the characteristics of LBA among modern mobile users, but also provide valuable references for the design, evaluation, and improvement of QoE-aware mobile applications and services.
\end{abstract}

\nop{

\ccsdesc[500]{Human-centered computing~Empirical studies in ubiquitous and mobile computing}
\ccsdesc[500]{Human-centered computing~Ubiquitous and mobile computing design and evaluation methods}
}


\keywords{low-battery anxiety, QoE quantification, video watching}

\settopmatter{printacmref=false}
\setcopyright{none}
\renewcommand\footnotetextcopyrightpermission[1]{}
\pagestyle{plain}

\maketitle

\thispagestyle{empty}

\section{Introduction}

``Have you ever ordered something at a bar just so you can ask to plug in your phone? Do you argue with loved ones because your phone died and you missed their calls or texts? Are you regularly accused of secretly `borrowing' someone else's charger? If so, you may be suffering from `Low-Battery Anxiety' '', according to a survey conducted by LG~\cite{rept_lowbatteryanxiety}. The survey also reported a shocking result\textemdash nine out of ten mobile users have the so-called low-battery anxiety (LBA), which refers to one's fear of losing mobile phone battery power especially when it is already at a low level ($20\%$ for example). The fear of losing battery power further triggers the ``no-mobile-phone phobia'' (also known as \textit{nomophobia}, refer to Sec.~\ref{subsec:lba-health} for the details), which is commonly considered as one social phobia and could lead to mental health problems~\cite{king2013nomophobia}. Considering the ever increasing number of smartphone users worldwide (surpassing three billions in 2019~\cite{statista2019}), the impact of accompanying LBA could be profound.


To accommodate the dying battery, people tend to change their behavior. For instance, the LG survey found that one in three people are likely to skip the gym, when it comes to choosing between hitting the gym and charging their smartphones. Those who severely suffer from the LBA were reported to behave strangely, e.g., head home immediately, ask chargers from strangers, secretly ``borrow'' other's charger, or stop answering friends' calls~\cite{rept_lowbatteryanxiety}. Also, the ubiquitous LBA is said to potentially harm our social relationships. Sixty percent of the LG surveyed mobile users blamed a dead phone for not speaking to their family members, friends or colleagues if their battery was low. \nop{Overall, as a consequence, the LBA could lead to negative effects on the mobile users' emotion, behavior and even health. Considering the anxiety triggered by a low-battery warning among billions of mobile users, we believe that the LBA is by no means a trivial issue any more.}

Understanding users' behavior when facing low battery level has a significant business meaning. In recent years, video streaming services at the mobile end are booming, triggered by the ubiquitous usage of mobile devices and emerging techniques in networking/computing~\cite{taleb2017multi}. Not only the already-popular over-the-top (OTT) video delivery, but also the new-generation HD, 4K or 8K video contents are making the mobile video streaming service one of the key features in the near 5G era. This new trend has fostered several fast-growing companies in mobile video-sharing services, such as Instagram and ByteDance. The view counts and the time of video watching directly impact the companies' revenue and customer retention rate. We have found that during the mobile video streaming, people tend to value between the attractiveness of a video and their battery power. This is mainly because playing video could consume a large portion (up to $55\%$) of the total power of a mobile phone~\cite{carroll2010analysis}. According to our findings from a large-scale user survey, mobile users may leave an attractive video at high probability when the battery power is low (Sec.~\ref{sec:behavior-quantify}). Therefore, for mobile video streaming service providers, LBA directly impacts the customer retention rate and should be regarded as an important quality of experience (QoE) metric.

A great deal of research has been devoted to saving energy and prolonging the battery lifetime of mobile phones. For example, tremendous efforts have been made to save energy of the major components of mobile phones, including CPU~\cite{yuan2003energy, kadjo2015control}, communication~\cite{lin2010energy, zhang2015etrain}, and display~\cite{stanley2016crayon, yan2017too}. Particularly in pervasive and ubiquitous computing, energy efficient techniques have been developed for mobile sensing/crowdsensing~\cite{priyantha2011littlerock,chon2016crowdsensing}, wireless communication~\cite{pyles2011sifi,ramos2011leap}, and battery management~\cite{banerjee2007users,min2015sandra}. Nevertheless, from the viewpoint of alleviating the LBA of mobile users, how effective were these energy saving approaches and how could we further improve them? These questions have remained largely unanswered. One of the biggest challenges to answer the above questions is the lack of a quantitative study of mobile users' LBA. As the anxiety of mobile users refers to a subjective feeling or emotion of human beings, it is hard to precisely quantify with some existing metric.

Generally speaking, ``if you cannot measure it, you cannot manage it.'' Due to the lack of LBA quantification, we can neither precisely measure the severity of anxiety among modern mobile users, nor accurately evaluate the effects of mobile energy saving strategies on LBA relief. A better understanding of mobile users' psychology and behavior leads to more effective and even new solutions to the LBA problem. This directly motivates our work in this paper: \textit{quantifying mobile users' LBA and as a case study investigating its impact on the mobile video streaming service.}

\nop{
Specifically speaking, we are also interested in mobile users' reactions to the LBA, for the following reasons. First, video streaming service at the mobile end is booming, triggered by the ubiquitous usage of mobile devices and emerging techniques in networking/computing~\cite{taleb2017multi}. Not only the already-popular over-the-top (OTT) video delivery, but also the new-generation HD, 4K or 8K video contents are making the mobile video streaming service one of the key features in the near 5G era. Second, the video playing could consume a large portion of the mobile system energy. During the video playing, the power consumption of the display component alone could account for $55\%$ of the total power of a mobile phone~\cite{carroll2010analysis}. Last but not least, the video streaming service providers are paying much attention to those factors impacting the audience retention, and as we will disclose (in Sec.~\ref{sec:behavior-quantify}), LBA is indeed one of them and may lead to a big customer loss. Thus, we aim to take the impact of LBA towards mobile video streaming service as a case study and quantify mobile user's reaction likelihood when experiencing different levels of LBA. 
}

Among existing ubiquitous computing research, however, very few work on LBA quantification can be found. Although there are relevant literature on the human-battery interaction (HBI)~\cite{heikkinen2012energy,rahmati2007understanding,banerjee2007users,ferreira2011understanding}, they are usually subject to two pitfalls: i) the analysis was performed in a qualitative way over the mobile users~\cite{rept_lowbatteryanxiety,ferreira2011understanding}, and thus the conclusions cannot be leveraged for quantitative evaluations; ii) the investigation was made for a specific and small user group~\cite{heikkinen2012energy,rahmati2007understanding,banerjee2007users}, and thus the obtained findings have limited generalizability for a large population. In this work, we not only conduct a large-scale investigation about the LBA issue, but also build LBA related quantitative models. Specifically, we set to answer three research questions that are critical and valuable to the interactive, mobile, wearable and ubiquitous technologies (IMWUT) community: 
\begin{itemize}
    \item (\textbf{RQ1}) How severe is the LBA among modern mobile users? 
    \item (\textbf{RQ2}) How can we quantify the LBA and extract quantitative models from the survey of a large-group of mobile users?
    \item (\textbf{RQ3}) How can we quantify the impacts of LBA on mobile users' video watching behavior, e.g., giving up watching an attractive video?
\end{itemize} 

By answering the above questions, we make the following contributions in this work.
\begin{itemize}
\item We conduct a user survey over $2000+$ mobile users and present a comprehensive study that addresses various aspects of the LBA issue. This leads to many interesting and insightful findings based on analysis of different user groups.
\item We present the methodology to extract the LBA curve through the surveyed data and obtain quantified anxiety degrees of the mobile users over varying battery levels. This provides a quantitative model for the mobile energy management research, especially the analysis and optimization of QoE-aware mobile applications/services.
\item We investigate the impact of LBA on mobile users' behavior, specifically their reaction to watching attractive videos. This leads to important quantitative findings on how the battery level is coupled with the mobile video watching behavior, and they thus provides valuable guidances for mobile video streaming services.
\item Based on our observations and findings through quantitative analysis and modeling, we give lessons and advice to mobile manufactures about HBI strategies and battery interfaces, to app developers/providers about energy-efficient mobile applications and mobile video streaming services.
\end{itemize}

The paper is organized as follows. We review the related work in Sec.~\ref{sec:relatedwork}. Sec.~\ref{sec:survey} presents the details of our survey. Sec.~\ref{sec:overall} gives an overview of the collected dataset and answers \textbf{RQ1}. We address \textbf{RQ2} and \textbf{RQ3} by quantifying the low-battery anxiety and its impacts on video watching behavior in Sec.~\ref{sec:lba-quantify} and Sec.~\ref{sec:behavior-quantify}, respectively. Further discussions are given in Sec.~\ref{sec:discuss} and the ethical statement is made in Sec.~\ref{sec:ethics}. Sec.~\ref{sec:conclude} concludes the paper.

\section{Related Work}\label{sec:relatedwork}

In this section, we first present the background knowledge of nomophobia, and then review relevant work on low-battery anxiety and range anxiety.

\subsection{Nomophobia, LBA and Mental Health}\label{subsec:lba-health}

Nomophobia, which is derived from ``No Mobile Phobia'', refers to the unreasonable fear of being unable to use a mobile phone or Internet. In the mental health domain, it belongs to a situational phobia related to the agoraphobia, along with fears of becoming ill and not receiving immediate assistance~\cite{king2014nomophobia}. With our lives increasingly relying on the mobile phone, nomophobia has been widely investigated and believed to have close relationship with our mental health~\cite{king2014nomophobia,sharma2015rising,tavolacci2015problematic,king2013nomophobia}. According to the findings in~\cite{king2014nomophobia}, nomophobia could increase people's anxiety, tachycardia, panic, fear and depression, especially for those who suffered from panic disorder and agoraphobia. The investigation in~\cite{sharma2015rising} showed that $83\%$ of the medical students under study experienced panic attacks when their mobile phones were misplaced. In~\cite{tavolacci2015problematic}, the authors found that over $30\%$ of the college students suffered from the anxiety caused by nomophobia and the women were affected more by the problematic use of mobile phones. It also presented in~\cite{king2013nomophobia} that nomophobic behavior could result in changes of daily habits or even mental disorders.

Nowadays, anxiety about the dying batteries (i.e., LBA) has become the major trigger for nomophobia~\cite{rept_diagnoselba}. Due to the aforementioned negative impacts of nomophobia, LBA may arouse greater concerns of modern people, especially when our younger generation starts to suffer from nomophobia~\cite{wiederhold2017digital}.

\subsection{LBA of Mobile Users}
Existing investigations can be found relevant to battery use and recharge behavior of mobile users, including \textit{qualitative} study in large-scale\footnote{The scale of study is debatable, since there is no broadly-accepted threshold counting for the large scale. We consider a study of sample size greater than $1,000$ people as large scale.}~\cite{rept_lowbatteryanxiety,ferreira2011understanding} or \textit{quantitative} study in small-scale~\cite{rahmati2007understanding,banerjee2007users,heikkinen2012energy}. Nevertheless, no much attention has been paid to the emerging LBA issue. Our work is the first to conduct a both \textit{large-scale} and \textit{quantitative} study specifically on LBA and its impacts.

\textbf{Large-scale qualitative studies:}
In 2016, LG conducted a survey among $2,000$ smartphone users in U.S., in which they found that: ``nine out of 10 people `felt panic' when their phone battery drops to 20 percent or lower''~\cite{rept_lowbatteryanxiety}. The phrase ``low-battery anxiety'' was firstly used in their survey report and caught a lot of attention. Thereafter, a successive relevant reports and news showed up in the press, especially recently. To name (the headlines of) some of them: ``The constant stress of the battery meter''~\cite{remove_battery_icon}, ``Diagnosing (and dealing with) your low-battery anxiety''~\cite{rept_diagnoselba}, and ``Low battery anxiety and how it relates to our mental health''~\cite{rept_lba_mental}. This indicates an awaking public concern towards the LBA issue.

Nevertheless, no quantitative analysis and models were disclosed, and there is no measure to quantify the anxiety degree and its impact. A study of battery charging behaviors with $4,035$ participants was conducted in~\cite{ferreira2011understanding}. The battery information of participants was collected via an Android application. Some statistical values were computed with the collected data, e.g., the average battery level, charging duration, charging schedule and frequency, based on which the authors obtained some charging behavior of participants, e.g., most users chose to interrupt the charging cycle which potentially reduces the battery life, and consistently overcharged their phones and tended to keep the battery levels above $30\%$. Nevertheless, neither quantitative models (beyond the statistical values) nor specific findings related to the LBA were provided.

\textbf{Small-scale quantitative studies:} 
In~\cite{rahmati2007understanding}, the authors conducted a survey among high school students and collected $350$ valid responses, among which only $41\%$ of the respondents were mobile users with average age of $17$ years old. According to the survey questionnaire, only qualitative self-estimation of LBA was performed. An small-group field study was also performed, in which the authors found that the mobile users could be categorized into two types regarding their charging behavior and they often have insufficient knowledge of the phone power characteristics. Although additional quantitative analysis on users' charging behavior was given, the field study was only made among $21$ mobile users. The focused small student group may limit the generality and accuracy of the conclusions drawn from the survey.
In~\cite{heikkinen2012energy}, both questionnaire studies and handset monitoring were performed, on users' attitudes to mobile devices' energy and their behavior on battery recharge. The studies involved up to $253$ participants, whose charging behavior was quantitatively analysed. The major findings include: most mobile users were good in estimating the energy consumption of mobile services, most mobile users were aware of power-saving settings and alters, and the mobile users demonstrated significant variation in battery use and recharge behavior. The anxiety caused by low-battery status, however, was not discussed in this work.
In~\cite{banerjee2007users}, the battery traces of $56$ laptops and $10$ mobile phones were collected and used to study the battery use and recharge behavior. The study yielded three major findings: i) users frequently recharged their devices even when the battery power level is not low, ii) the charging behavior was driven by context or battery levels, and iii) there were significant variations in patterns shown by the users. Based on the user study, the authors designed a battery energy management system, called Llama, to enhance existing energy management policies. Nevertheless, the number of participants is too small to draw a reliable conclusion with respect to LBA. \nop{ small-scale user study, the obtained findings may not be unreliable to support their system design. }

\nop{
\subsection{Range Anxiety of Battery Electric Vehicle Drivers}

Battery electric vehicle (BEV) drivers are exposed to the so-called range anxiety, i.e., the fear of being left in the middle of a trip due to a BEV's empty battery. It has been investigated for long in the transportation and power system fields~\cite{neubauer2014impact,rauh2015understanding,franke2012experiencing}. Although range anxiety and LBA seem similar, the results and conclusions for the range anxiety cannot be directly applied to the case of LBA. First, the availability of charging sources for mobile phones and for BEVs is completely different. Such a difference will certainly lead to different charging behavior. Second, mobile phones have a much higher market penetration rate than BEVs. As such, people are more dependent on the mobile phones for their daily lives and may care more about the phone battery life. As a result, the impacts of battery level of mobile phone on users may be more significant than the range anxiety of BEVs, observed by the much more frequently charging behavior of mobile users~\cite{heikkinen2012energy,rahmati2007understanding,banerjee2007users}. Last but not least, drivers can avoid range anxiety by driving gasoline vehicles, but mobile users are not granted for such an option since all mobile phones are battery powered.
}

\section{A Survey over $2000+$ Mobile Users}\label{sec:survey}

To investigate the severity of LBA and quantify its impacts on mobile users, we i) carefully designed a well-informed questionnaire, ii) extensively distributed it over a popular mobile crowdsouring (MCS) platform, and iii) continuously conducted the survey for over three months. At the end, we collected feedback from $2,071$ mobile users.

\subsection{Questionnaire Design and Distribution}
It is not easy to conduct a large-scale user survey for the LBA study. \textbf{First}, the questionnaire designed for large-scale investigation should be as short as possible~\cite{questionnaire_siap}, because people may simply not participate if the survey takes too long. \textbf{Second}, the design of questionnaire is tricky. On the one hand, complex (technical) questions directly asking about the anxiety could lead to inaccurate and unreliable feedback, as self-estimation of LBA is highly subjective and the participant may not understand the questions well~\cite{heikkinen2012energy}. On the other hand, with simple questions we may not be able to obtain sufficient information (to extract quantitative models) from participants. \textbf{Third}, it may be costly for distributing the questionnaire and soliciting answers in large scale. To distribute hundreds or thousands of questionnaires is a heavy task, let alone to have them filled out and fetch them back. \nop{An easy way to complete this task is to seek help from the social network, while the platform chosen to show the questionnaire and interact with the users could make big differences.}

\begin{figure}[!t]
\begin{center}
\includegraphics[width=0.7\textwidth]{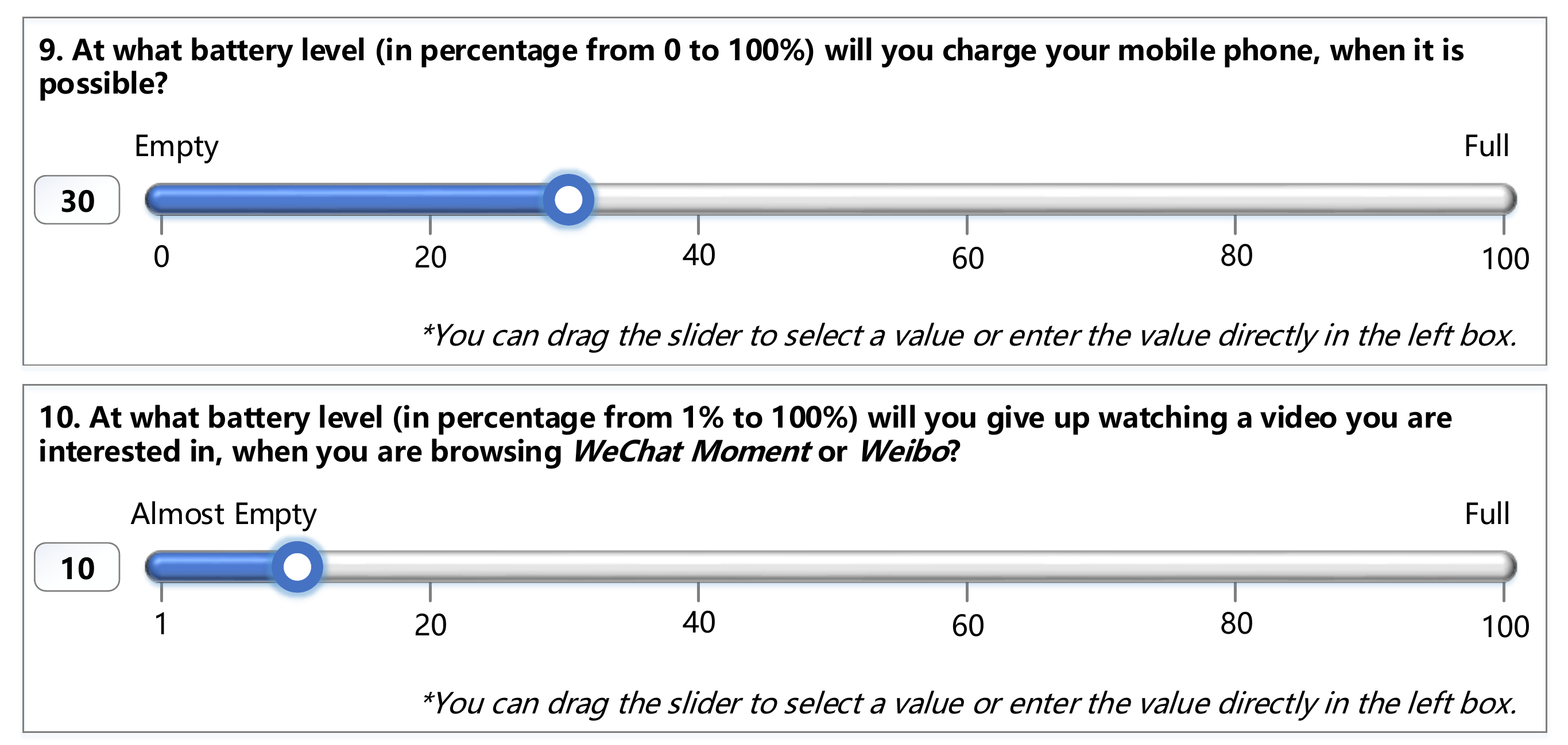}
\caption{The $9$th and $10$th questions applying the sliding strip style in our questionnaire.}\label{fig:sliding_strip}
\end{center}
\end{figure}

To tackle the first two difficulties, we design our questionnaire with the following principles:
\begin{itemize}
\item \textbf{Short questionnaire}: By precisely defining our purpose of this investigation and carefully setting tailored questions (through a survey validation process), we manage to cut down the questionnaire to ten questions. Refer to Appendix~\ref{sec:apdx_survey} for the detailed questionnaire. This directly cuts down the time needed for completing the questionnaire. According to our statistic data, on average it took $88.4$ seconds for a participant to finish the survey, and over half of the participants finished it within one minute.

\item \textbf{Elaborate UI}: Choice questions are adopted most in the traditional questionnaire design. Nevertheless, such question style is limited in obtaining digitalized feedback with many choices. In our case for example, for the question of ``\textit{at what battery level will you charge your mobile phone, when it is possible?}'', it is cumbersome to set up a large number of choices for a fine-grained feedback, if not impossible. To obtain quantified (digital) feedback from participants in a more convenient and friendly way, we used the question style of \emph{sliding strip/bar}, as illustrated in Fig.~\ref{fig:sliding_strip}. The participant can slide the bar freely to choose what she/he wants, or directly fill a desired value in the left box. With the very intuitive user interface, the participants need zero learning time and can easily answer the questions in any specified resolution (e.g., $1\%$ in our setting).

\item \textbf{Avoid difficult questions}: The straightforward question for LBA would be ``at what battery level will you experience low-battery anxiety?'' Nevertheless, it is not easy to answer this question because different people may have a different understanding about the vague term ``anxiety''. To obtain reliable answers, we instead designed easy-to-understand and easy-to-answer questions, for example, ``at what battery level (in percentage from $0$ to $100\%$) will you charge your mobile phone, when it is possible?'' and ``at what battery level (in percentage from $1\%$ to $100\%$) will you give up watching a video you are interested in, when you are browsing the WeChat Moment or Weibo?'' These questions can be easily answered based on users' own daily experience. Together with another easy-to-answer question ``when inconvenient to charge the mobile phone, will you experience anxiety or panic when the battery level is low?'', we can obtain more reliable answers for quantifying the LBA models. 
\end{itemize}

To tackle the third difficulty, we resort to the help from social networks and distribute our survey over a popular MCS platform.  
\begin{itemize}
\item \textbf{Popular MCS platform}: With over one billion users, \emph{WeChat} is currently the most popular social network and instant messaging mobile app~\cite{wechat_popularity}. It is regarded as an ideal platform for MCS tasks in our case. Aided by the \emph{WeChat Mini Program}~\cite{wechat_miniprogram}, it is easy to create a questionnaire meeting all our requirements. Furthermore, it greatly reduces our efforts to distribute the questionnaire to large-scale users, with WeChat's inherent functions of message sharing and group chatting\footnote{A chatting group in WeChat is a gathering of multiple users (from $3$ to $500$) with/without friend relationship. It is most convenient for information sharing (e.g., the questionnaire distribution in our case) in a group of people and widely used among WeChat users.}.
\end{itemize}

Note that although the questions raised in the questionnaire look simple, they were by no means from straightforward intuitions or random guesses. Some of them went through a reverse-engineering alike process by asking how to implicitly derive the anxious degree under various battery status. We demonstrate that, with such settings, simple questions can become powerful enough to extract quantified LBA model over the massive collected answers (e.g., with a reversed accumulative operation in Sec.~\ref{sec:lba-quantify}).

\subsection{Data Cleansing}
\nop{Kui: Very dangerous here. Using incentive may bias the survey results. It simply opens a jar of worms that we cannot defend! Double check if other part of the paper has the similar problem. As an incentive to participate the survey, the participants had a chance of winning a prize in a prize draw at the end of the survey. The prizes were sponsored by companies posting advertisements over WeChat.}

The raw data collected by the MSC platform also include some meta information of each answer, including: 1) submission time, 2) time spent (to complete the questionnaire), and 3) mobile phone IP address and approximate location (at city level). The raw data (answers) from all participants looked quite cleaning on the surface, as there were no missing, corrupted or ill-formatted entries. To further clean the data and wipe out those reckless and unreliable answers, we exploit the semantic consistency lying in the reliable answers. Specifically, we design the following methods to clean up the data:
\begin{itemize}
    \item \textbf{Reckless answers:} With the meta information of \textit{time spent}, we filtered out the answers with too short time (less than $30$ seconds in our case), which occupy around $1\%$ of the total answers.
    \item \textbf{Unreliable answers:} We applied association rules to filtered out contradictory or unreasonable answers. i) The ``student'' with age ``older than $65$'' were filtered out (reason: it is unlikely while not impossible); ii) those ``severely suffering'' from the LBA while with charging threshold below $10\%$ were filtered out (reason: too obvious); iii) those who claimed ``satisfied'' with their batteries while with necessary charging over 4 times per day were filtered out (reason: contradicting).
\end{itemize}

After the data cleansing, we eventually obtained $2,032$ effective answers.

\section{Dataset and Preliminary Analysis (RQ1)}\label{sec:overall}

\subsection{About Participants and Mobile Phones (Table~\ref{tbl:participants})}

\definecolor{Gray}{gray}{0.9}
\definecolor{LightCyan}{rgb}{0.88,1,1}
\setlength{\arrayrulewidth}{0.5mm}

\begin{table}[tb]
\centering
\caption{Survey subjects and corresponding frequencies (participants and mobile phones, $N=2,032$).}
\label{tbl:participants}
\begin{tabular}{l l}
{Survey Subjects} & {Frequency (\%)} \\
\hline
\rowcolor{Gray}
{Meta Info.} &  \\
\# Cities & 150  \\
\# Provinces* & 31 \\
\# Countries & 11 \\
\rowcolor{Gray}
{Gender (Q1)} &  \\
Male &  1095 (53.89)  \\
Female & 937 (46.11) \\
\rowcolor{Gray}
{Age (Q2)} & \\
Under $18$ & 9 (0.52)\\
$18\sim 25$ & 888 (51.45) \\
$25\sim 35$ & 460 (26.65) \\
$35\sim 45$ & 250 (14.48) \\
$45\sim 65$ & 119 (6.89) \\
\rowcolor{Gray}
{Occupation (Q3)} & \\
Student & 1024 (50.39)\\
Gov/Inst & 271 (13.34) \\
Company & 434 (21.36) \\
Freelance & 144 (7.09) \\
Others & 159 (7.82) \\
\rowcolor{Gray}
{Mobile Phone Share (Q4)} &  \\
iPhone & 737 (36.27)  \\
Huawei & 682 (33.56)\\
Xiaomi & 228 (11.22) \\
Others & 385 (18.95)\\
\hline
\multicolumn{2}{r}{\textit{*Provinces refer to provincial-level administrative units of China.}}
\end{tabular}
\end{table}

\begin{itemize}
\item \textbf{Geo-distribution of participants}: According to the meta information, the majority of participants came from $150$ cities in $31$ provincial-level administrative units of China. Meanwhile, we also noticed a small number ($70$ out of $2,032$) of oversea participants, which were mainly from America, Canada, Singapore, Australia, and Japan.
\item \textbf{Gender (Q1)}: Males account for $53.89\%$ and females account for $46.11\%$ of the total participants, resulting in a generally balanced gender distribution.
\item \textbf{Age (Q2)}: This question was not mandatory and $1,726$ participants answered it. From the feedback, most ($78.62\%$) of the participants are young people, with $51.45\%$ aged between 18 and 25, and $26.65\%$ aged between 25 and 35. Participants with age above $35$ occupy about $20\%$.
\item \textbf{Occupation (Q3)}: Most ($50.39\%$) participants are college students, including both undergraduates and post-graduates; participants from companies, institutes and governments occupy $34.70\%$.
\item \textbf{Mobile phone share (Q4)}: iPhone and Huawei hold the largest mobile phone shares in our investigation, accounting for $36.27\%$ and $33.56\%$, respectively. Xiaomi follows them with a share of $11.22\%$, and other brands (Oppo, Vivo, Samsung, etc.) contribute to the rest $18.95\%$ share. \nop{comment: is this consistent with the reported mobile phone market share in China?}
\end{itemize}

Overall, we can see that the majority of participants in our survey are Chinese college students, with age between 18 and 35. The most popular mobile phone brands among them are iPhone and Huawei.

\subsection{User Satisfaction of Mobile Phone Battery (Table~\ref{tbl:battery})}

\begin{table}[tb]
\centering
\caption{Survey subjects and corresponding frequencies (battery capacity status, $N=2,032$).}
\label{tbl:battery}
\begin{tabular}{l l}
{Survey Subjects} & {Frequency (\%)} \\
\hline
\rowcolor{Gray}
{Battery Satisfaction (Q5)} &  \\
Satisfied & 819 (40.31)  \\
Just OK & 836 (41.14) \\
Not Satisfied & 377 (18.55) \\
\rowcolor{Gray}
{Necessary Charging Frequency (Q6)} &  \\
$\leq 1$ & 764 (37.60)  \\
$2$ &  1002 (49.31) \\
$3$ & 191 (9.40) \\
$\geq 4$ & 75 (3.69) \\
\rowcolor{Gray}
{Portable Charging Frequency (Q7)} & \\
Never Use & 400 (19.69)\\
Occasionally Use & 1423 (70.03) \\
Frequently Use & 209 (10.29) \\
\hline
\end{tabular}
\end{table}

\begin{itemize}
\item \textbf{Battery satisfaction (Q5)}: $40.31\%$ of the participants were satisfied with the battery capacity of their mobile phones, while $18.55\%$ were not. The rest participants ($41.14\%$) were \textit{just OK} with their battery capacities.
\item \textbf{Necessary charging\footnote{For necessary charging, it means that the user has to charge the mobile phone, as it will run out of power soon.} frequency (Q6)}: $37.60\%$ of the participants charged their mobile phones once (or less) a day. Nearly half ($49.31\%$) of the participants charged their mobile phones twice a day, while about $13\%$ charged mobile phones more than twice per day.
\item \textbf{Portable charging frequency (Q7)}: Over $80\%$ of the participants used portable charging (backup battery) for mobile phones, with $70.03\%$ occasional usage and $10.29\%$ frequent usage.
\end{itemize}

Based on the above feedback, we can see that there is still room to further improve user satisfaction of current battery capacity design.

\subsection{Severity of Low-Battery Anxiety (Table~\ref{tbl:severity})}

\begin{table}[ht]
\centering
\caption{Survey subjects and corresponding frequencies (severity of LBA suffering, $N=2,032$).}
\label{tbl:severity}
\begin{tabular}{l l}
{Survey Subjects} & {Frequency (\%)} \\
\hline
\rowcolor{Gray}
{Suffering of LBA and its Severity (Q8)} &  \\
Not a Little & 165 (8.12)  \\
A Little & 1173 (57.73) \\
Confirmed Suffering & 570 (28.05) \\
Severely Suffering  & 124 (6.10) \\
\hline
\end{tabular}
\end{table}

Given the definition and ``symptoms'' of LBA, for the eighth question (Q8), the participants were required to self-estimate the severity of LBA they were suffering, especially when they were under inconvenient charging situations. Based on the feedback, it is surprising that $91.88\%$ of the participants are suffering from the LBA, more or less. This is consistent with LG's survey~\cite{rept_lowbatteryanxiety}, but the percentage is even higher in ours. Particularly, over $34\%$ of them firmly admitted their suffering of LBA, with $6.10\%$ of which claimed ``severely suffering''.

To this end, we can almost conclude that the LBA among mobile users (at least the $2,032$ participants) are severe. It could also be confirmed with the feedback of Q6 and Q7. In the next section, we conduct a more thorough investigation on the LBA. 

\nop{Although severe, how severe (i.e., to what extent) of the LBA is what we are interested and need to explore in the following section.}

\section{Quantification of Low-Battery Anxiety (RQ2)}\label{sec:lba-quantify}

To quantify the LBA of mobile users is tricky. It is (methodologically) infeasible and (technically) inaccurate to ask a user to provide a real value representing her/his anxiety degree at each battery level. Thus, in our user survey, instead of directly asking for the anxiety degree, we turn to ask the battery level at which the user will charge the mobile phone, and then extract an LBA curve from the feedback of all participants.

\subsection{Extraction of LBA Curve}

Specifically, the ninth question (Q9) of our questionnaire is set as: \textit{at what battery level (in percentage from $0$ to $100\%$) will you charge your mobile phone, when it is possible?} The answer provides us with an angle to infer at which energy level the user begins to worry about the battery life, i.e., experience the low-battery anxiety. With all the answers from the $2,032$ participants, we are able to extract an empirical \emph{LBA curve} by reversely accumulating over the histogram (showing the frequency at which users begin to experience the low-battery anxiety) and normalizing the cumulative numbers to $[0, 1]$. The detailed procedure is as follows:
\begin{enumerate}
\item \textbf{Initializing}: We first set $100$ empty bins, labelled from $1$ to $100$, indicating the battery levels from (almost) empty to full;
\item \textbf{Counting}: For each of the answer, e.g., $a$ (an integer in the region of $[1, 100]$), we add one to each of the bins with labels in $[1, a]$;
\item \textbf{Cumulating}: We conduct the above operation for all answers and get the $100$ bins with cumulative numbers, resulting in a declined (discrete) curve in the region of $[1, 100]$;
\item \textbf{Normalizing}: By normalizing the cumulative numbers to $[0, 1]$ to represent the anxiety degree, we obtain the LBA curve: \emph{battery level} vs. \emph{anxiety degree}.
\end{enumerate}

The pseudocode of the above process is given in Algorithm~\ref{alg:LBAextract} (in the \textit{Python} programming style). Note that although the LBA degree extraction process seems simple, it is the result of a reverse-engineering idea and corresponding questionnaire design. Specifically, to learn the anxious degree curve of the mobile user with just one question, the question requires to make indications across different battery levels. By implicitly asking when a user begins to feel anxious about her/his battery power (i.e., the charging threshold), we are able to harvest the anxiety indications across all battery levels of the user, theoretically by a 0-1 binary vector with 0 indicating ``not anxious'' and 1 the opposite. Then, using a reversed accumulation approach among the 2000+ collected answers (distinguished from the standard histogram plotting approach), the empirical LBA curve (anxious degrees vs. battery levels) can thus be derived.

\begin{algorithm}[!tb]
\caption{LBA Degree Extraction (Reversed Accumulation of the Charging Threshold Distribution)}\label{alg:LBAextract}
\KwIn {$inputs\_Q9$, the vector consisting of Q9 answers of all $N$ participants.}
\KwOut{$anxiety\_degree$, the vector of anxiety degrees, corresponding to battery levels from $1\%$ to $100\%$.}

\textcolor{MyBlue}{\# \texttt{(1) Initializing}\\}
$counter = [0 \text{ \textbf{for} } i \text{ in range}(100)]$\\

\textcolor{MyBlue}{\# \texttt{(2) \& (3): Counting \& Cumulating}\\}
\For{$value \emph{ in }inputs\_Q9$}
{
	\For{$j \emph{ in range}(value)$}
	{
		$counter[j] = counter[j]+1$ \\
	}
}

\textcolor{MyBlue}{\# \texttt{(4) Normalizing}\\}
$anxiety\_degree = [count/N \text{ \textbf{for} } count \text{ in } counter]$ \\

return $anxiety\_degree$
\end{algorithm}

\subsection{Observations and Analysis}

\subsubsection{Overall LBA Curve}

\begin{figure}[!h]
\begin{center}
\includegraphics[width=0.55\textwidth]{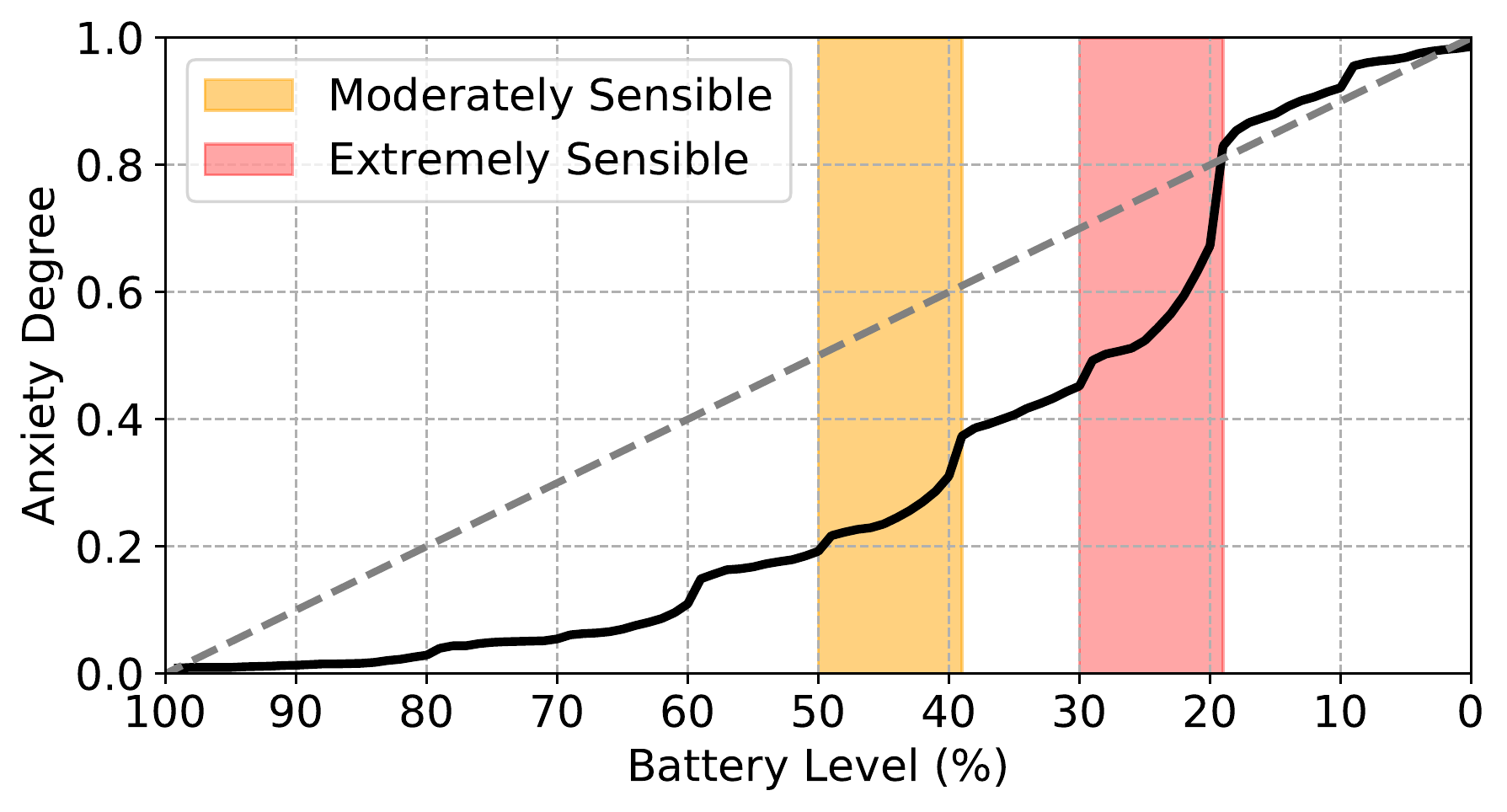}
\caption{Extracted LBA curve (the black solid line) from the survey data of $2,032$ mobile users, where the anxiety sensible regions are highlighted in orange and red.}\label{fig:anxiety_overall}
\end{center}
\end{figure}

The extracted LBA curve for all the participants is shown in Fig.~\ref{fig:anxiety_overall}, and the detailed values of anxiety degrees under different battery levels are listed in the Appendix. From the resulted LBA curve, we can observe that:

\begin{itemize}
\item The anxiety degree does not linearly increase with the decrease of battery level. As a comparison, we draw a straight line in Fig.~\ref{fig:anxiety_overall} (i.e., the grey dashed line), indicating the linear increase situation. It can be found that, the LBA curve is (approximately) convex to the battery level in $[20\%, 100\%]$, while is (approximately) concave when the battery level drops to $[0, 20\%]$. This observation shows that, the mobile user gets more sensible to the battery level as the energy drains.\nop{ until the battery level drops below a threshold ($20\%$ from our observation).}
\item Two sensible regions of the users' anxiety can be found, named \emph{moderately sensible} and \emph{extremely sensible} regions as illustrated in Fig.~\ref{fig:anxiety_overall}, both corresponding to about $10\%$ battery level drop. In the moderately sensible region, the $10\%$ battery level dropping leads to $18\%$ anxiety degree increases (from $0.19$ to $0.37$), while in the extremely sensible region the number is $38\%$ (from $0.45$ to $0.83$). The occur of the extremely sensible region is most probably due to the battery user interface (e.g., the battery icon's color changes to yellow or red) and the low-battery warning message from the mobile OS.
\end{itemize}

\subsubsection{LBA Curves for Different Age Groups}

\begin{figure}[!h]
\begin{center}
\includegraphics[width=0.55\textwidth]{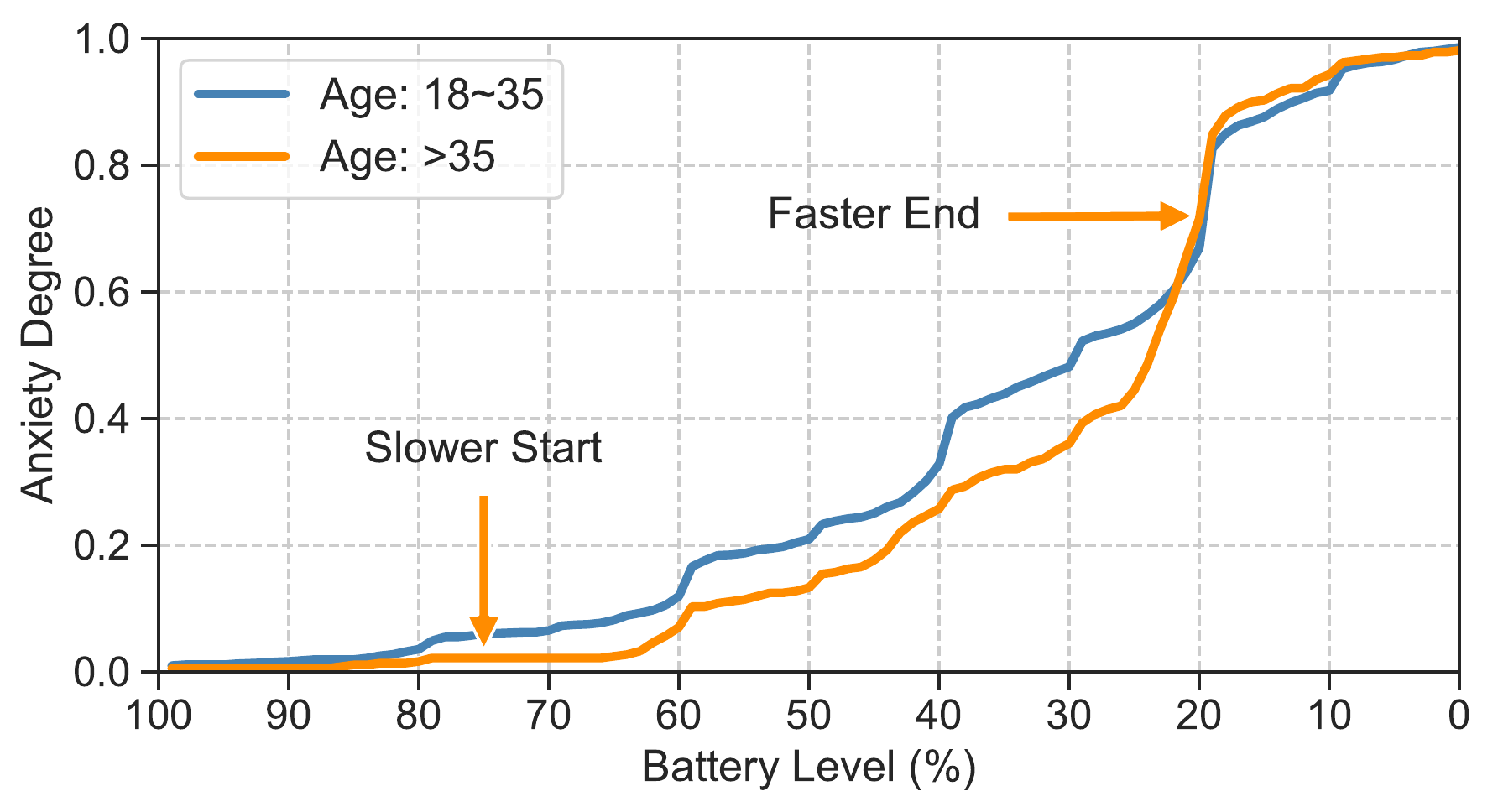}
\caption{LBA curves for different user age groups. The ``slower start'' and ``faster end'' characters can be found in the anxiety curve of the user group older than 35, compared with the younger users aged between $18$ and $35$.}\label{fig:anxiety_age}
\end{center}
\end{figure}

We further partition the participants into two groups: younger users aged between $18$ and $35$, and older users aged above $35$. We then extract the LBA curves for the two groups respectively. The results are illustrated in Fig.~\ref{fig:anxiety_age}, from which we find that:
\begin{itemize}
\item When battery levels are in regions of $[25\%, 40\%], [45\%, 60\%]$ and $[65\%, 90\%]$, the younger users are significantly more anxious about their mobile phones' battery levels than the older users (\textcolor{MyBlue}{$all$ $p < 0.05$})\footnote{A $P \leq 0.05$ is considered as statistically significant for the purpose of this study.}. This makes sense, as nowadays the younger users spend more time and thus are more dependent on the mobile phone than the older users~\cite{rept_timespentonphone}.

\item When the battery level is low (e.g., in the region of $(0, 20\%]$), the older users start getting more sensible to the dropping of battery level, and their anxiety goes up rapidly and becomes higher than that of the younger users. The overall anxiety of the older users shows ``slower start'' and ``faster end'' phenomena, compared with that of the younger users. \nop{This is an interesting phenomenon needing a further investigation.}
\end{itemize}

\subsubsection{Role of Gender in Battery Charging}

\begin{figure}[!h]
\begin{center}
\includegraphics[width=0.75\textwidth]{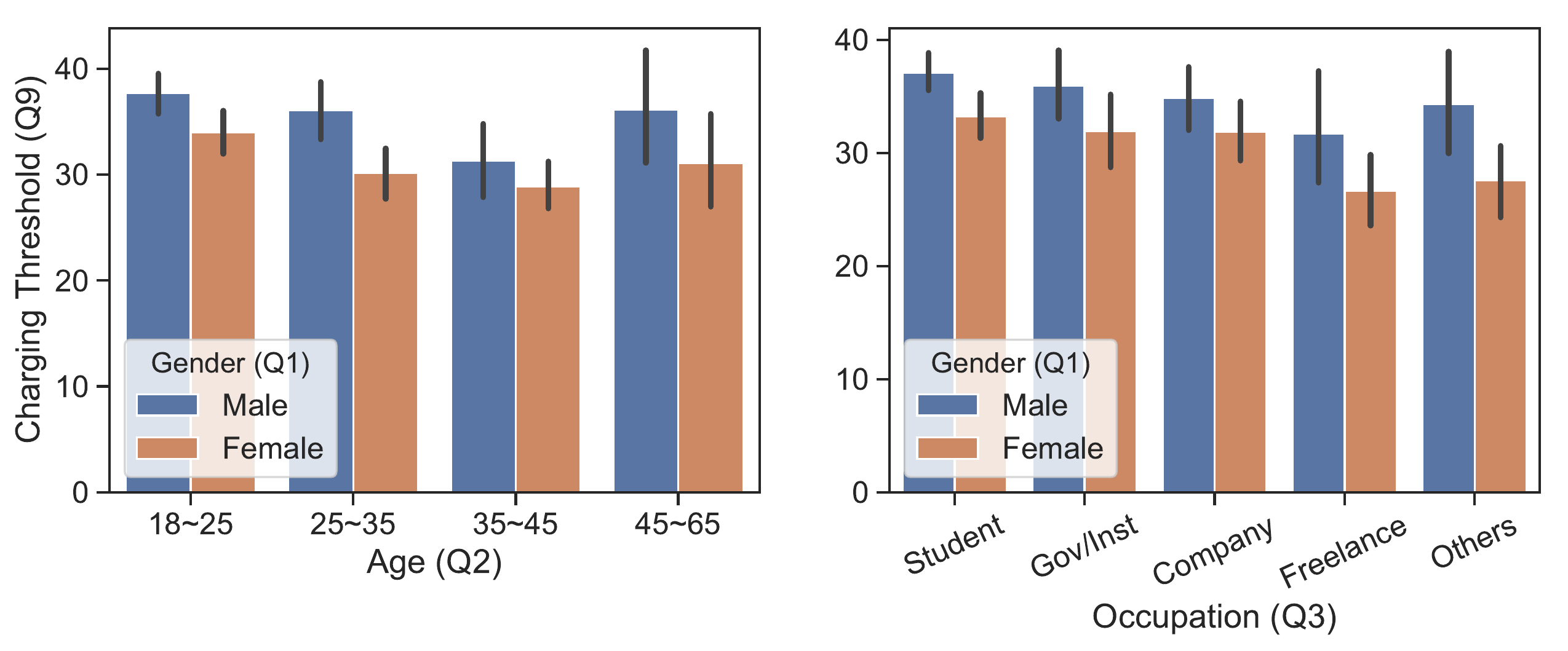}
\caption{Role of gender on the battery charging threshold, for user groups with different ages and occupations, respectively, from which we can find that females charge their mobile phones significantly earlier than males (\textcolor{MyBlue}{$p=0.05$}).}\label{fig:anxiety_gender}
\end{center}
\end{figure}

In addition to the overall and age-respective analysis of the LBA curve, we also explore the roles of different human features in battery charging. One apparent finding is about the difference brought by gender. With the results illustrated in Fig.~\ref{fig:anxiety_gender}, we find that females charge their mobile phones (i.e., get anxious about the battery level) significantly earlier than males (\textcolor{MyBlue}{$p=0.05$}), regardless of age and occupation. A similar conclusion was given in~\cite{tavolacci2015problematic}, where females had a higher score associated with the (negative) impact of nomophobia due to the problematic use of mobile phones.

\subsection{Lessons Learnt from LBA Quantification}

\begin{itemize}
\item To the developer/provider of energy-efficient or low-power applications/services: The LBA curve could be utilized to improve users' quality of experience (QoE), and the users staying in sensible regions could be the potential targets, compared to those in insensible regions. For example, in the work of~\cite{zhang2019e2e}, the authors adopted a similar idea to improve the user QoE of web service, by identifying and utilizing users' most sensible region for the webpage loading delay.
\item To the mobile phone manufactures and mobile OS developers: During the designs of HBI strategy and battery interface~\cite{rahmati2007understanding}, they should pay attention to the anxiety sensible regions, especially the extremely sensible one. Some measures, e.g., more efficient energy management strategies~\cite{banerjee2007users} and considerate GUI designs~\cite{vallerio2006energy}, could be taken to smooth the curve (thus soothe the anxiety) in these regions. Furthermore, the younger mobile users may need more attention mainly due to their continuous higher anxiety over the battery lifetime. Some adaptations of the battery interface could be made, e.g., releasing sufficient information about the remaining battery energy~\cite{abrahamse2007effect} or providing accurate and personalized battery lifetime predictions~\cite{kang2011personalized,oliver2011empirical}, to alleviate the anxiety and improve QoE.
\end{itemize}

\section{Impacts of LBA on Video Watching (RQ3)}\label{sec:behavior-quantify}

\nop{
It has been reported that video could consume a large portion of the mobile system energy~\cite{trestian2012energy}. During the video playing, the power consumption of the display component alone could account for $55\%$ of the total power of a mobile phone~\cite{carroll2010analysis}. Thus, we are motivated to investigate and quantify the impact of LBA on mobile users' video watching behavior, specifically at which battery level the users will \emph{give up} watching (attractive) videos.
}

In this section, we investigate the impacts of LBA on mobile user's behavior, and specifically we are interested to see the mobile user's reaction in watching videos (i.e., during video streaming) when experiencing different levels of low-battery anxiety. 

\nop{
\subsection{Why Video Streaming?}

There could be more interesting or meaningful applications or services for the investigation, such as navigation, tickets booking/displaying, etc. Nevertheless, the reactions of mobile users towards LBA in those circumstances are not easy to quantify (if not impossible), as the users usually have no choice but endure the LBA, e.g., a driver in a new city most probably will not stop the navigation service just because the battery power is low. For video streaming, however, the audiences could decide more freely to leave or stay depending on their values of the battery power. This gives us an opportunity to measure the possibility (or likelihood) how the users behave under different battery status. Thus, we are motivated to investigate and quantify the impact of LBA on mobile users' video watching behavior, specifically at which battery level the users will \emph{give up} watching (attractive) videos.
}

\subsection{Extraction of Video Abandoning Likelihood Curve}

In the last question (Q10) of our questionnaire, we ask the participants to answer: \textit{at what battery level (in percentage from $1\%$ to $100\%$) will you give up watching a video you are interested in, when you are browsing the WeChat Moment or Weibo\footnote{The \emph{WeChat Moment} and \emph{Weibo} are currently the two biggest mobile social network platforms in China with billions of users. Tons of fresh and popular videos are shared there and updated by second.}?} The feedback to this question directly shows how the mobile users value the battery power versus attractive videos, thus indicates how the LBA impacts the behavior of video watching.

Following a similar way of extracting the LBA curve (i.e., Algorithm~\ref{alg:LBAextract}, with $inputs\_Q9$ replaced by $inputs\_Q10$), we were able to extract a curve (illustrated in Fig.~\ref{fig:videoabandon_overall}) called \emph{video abandoning likelihood}. The curve actually indicates the likelihood\footnote{Note that this is not probability, as the integral below the curve is not equal to one.} that a user may abandon video watching under different battery levels. From another perspective, the curve depicts the (approximate) proportion of mobile user, among the whole population, that will give up watching attractive videos at specific battery levels.

\subsection{Observations and Analysis}\label{subsec:observe_videoabandon}

\subsubsection{Overall Video Abandoning Likelihood}

\begin{figure}[!ht]
\begin{center}
\includegraphics[width=0.55\textwidth]{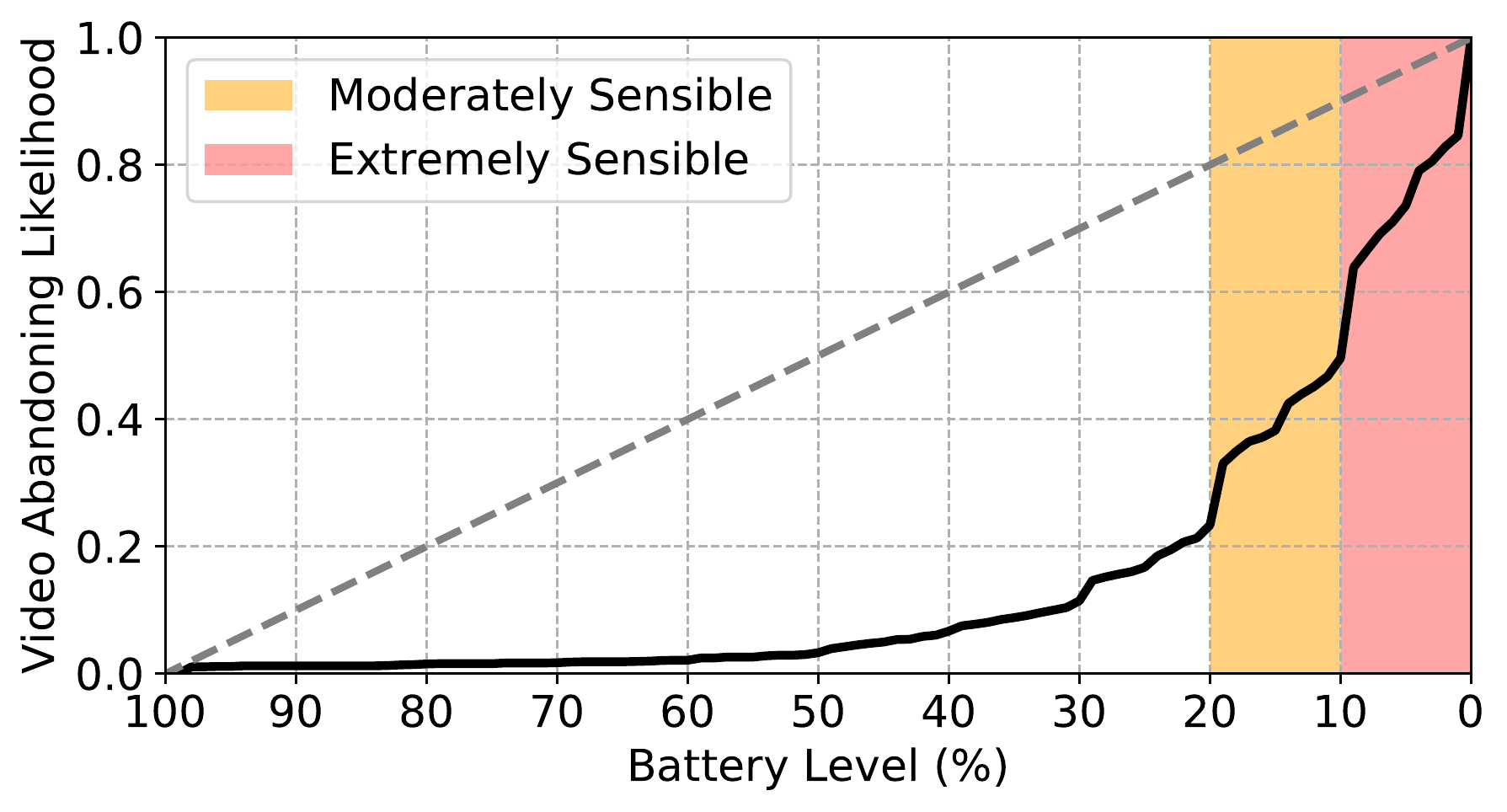}
\caption{The extracted video abandoning likelihood curve (the black solid line) from the survey data of $2,032$ mobile users. The video abandoning sensible regions are highlighted in orange and red.}\label{fig:videoabandon_overall}
\end{center}
\end{figure}

The detailed values of video abandoning likelihood under different battery levels are provided in the Appendix. According to the video abandoning likelihood curve in Fig.~\ref{fig:videoabandon_overall}, we have the following observations.
\begin{itemize}
\item The video abandoning likelihood does not linearly increase with the draining of battery power. In contrast, the curve is way below the linear trend (the grey dashed line in the figure), indicating that mobile users generally value attractive videos much more than the battery energy.
\item When the battery level is above $20\%$, the user's video watching behavior seems not much affected by the battery level, with a video abandoning likelihood less than $0.23$. Nevertheless, when the battery level drops below $20\%$, the video abandoning likelihood rises up quickly. Specifically, when the battery energy is left around $10\%$, nearly half ($49.60\%$) of the mobile users will give up watching (attractive) videos.
\item Two sensible regions of the video abandoning likelihood curve can be found: the \emph{moderately sensible region} for battery level in $[10\%, 20\%]$, corresponding to an abandoning likelihood increase of $0.26$; the \emph{extremely sensible region} for battery level in $(0, 10\%]$, corresponding to an abandoning likelihood increase of $0.50$.
\end{itemize}

\subsubsection{Video Abandoning Likelihood for Different Age Groups}

\begin{figure}[!h]
\begin{center}
\includegraphics[width=0.55\textwidth]{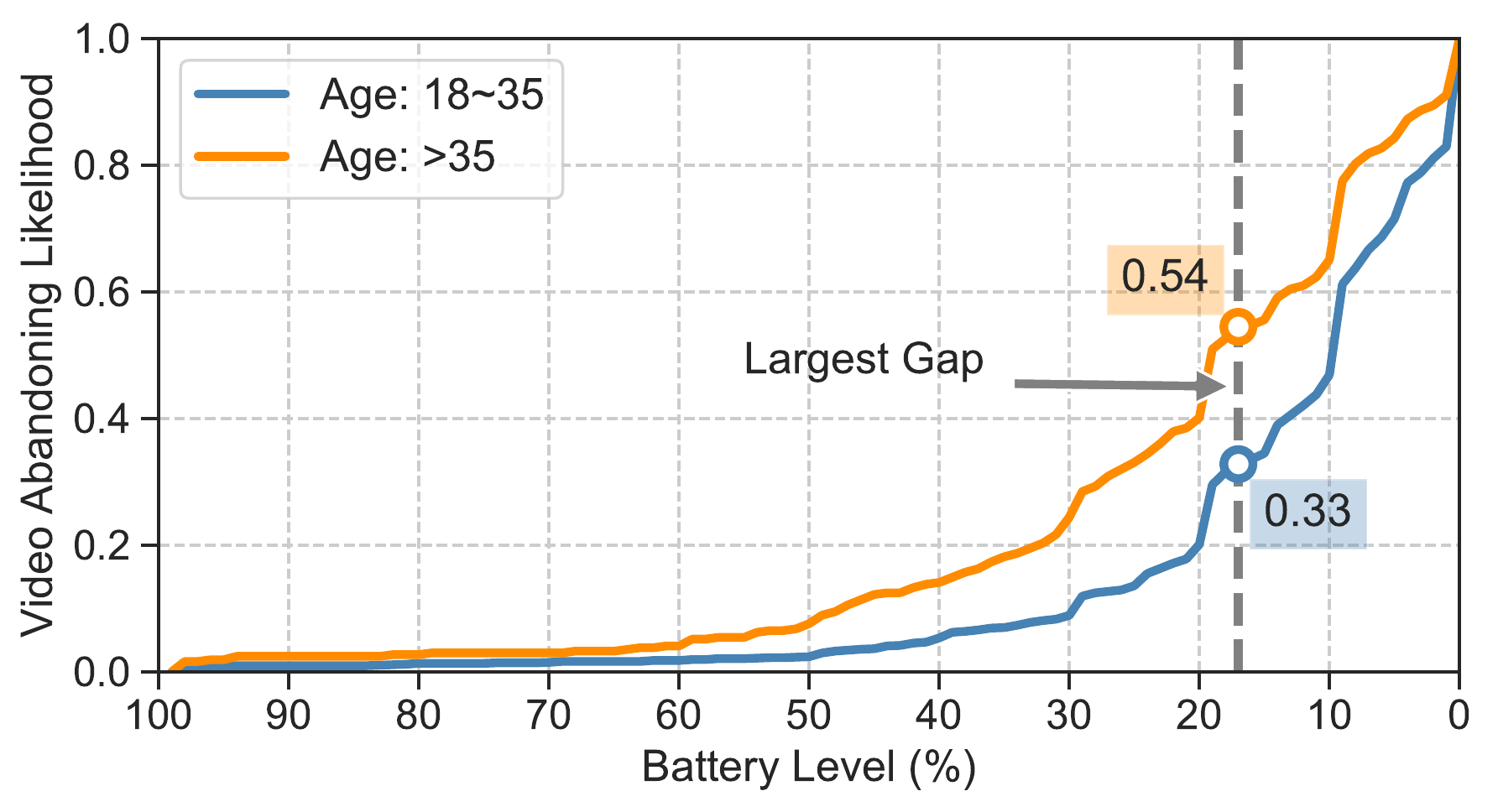}
\caption{The video abandoning likelihood under different age groups.}\label{fig:videoabandon_age}
\end{center}
\end{figure}

We further partition the users into two groups: younger users with ages between $18$ and $35$, and older users with age above $35$. Then, we extract the video abandoning likelihood curves for both groups. The results are illustrated in Fig.~\ref{fig:videoabandon_age}, from which we can find:
\begin{itemize}
\item The older users value the battery power more than the younger users. The video abandoning likelihood of older users is larger than that of the younger users, significantly when the battery level is low ($[10\%, 30\%]$, \textcolor{MyBlue}{$p=0.05$}). This copes with the fact that younger users spend more time on mobile apps (including video watching) than older users~\cite{rept_timespentonapps}.
\item Specifically, the largest gap between the two curves appears at the battery level of $17\%$, with video abandoning likelihoods of $0.54$ and $0.33$, respectively. This indicates that, when the battery energy drops to $17\%$, over $60\%$ more users aged above $35$ than those between $18$ and $35$ will give up watching attractive videos.
\end{itemize}

\subsubsection{Correlation between Battery Charging and Video Abandoning}

\begin{figure}[!h]
 	\centering
 	\subfigure[Overall regression plot.]{
 		\label{fig:videoabandon_corr_reg}
 		\includegraphics[width=0.45\textwidth]{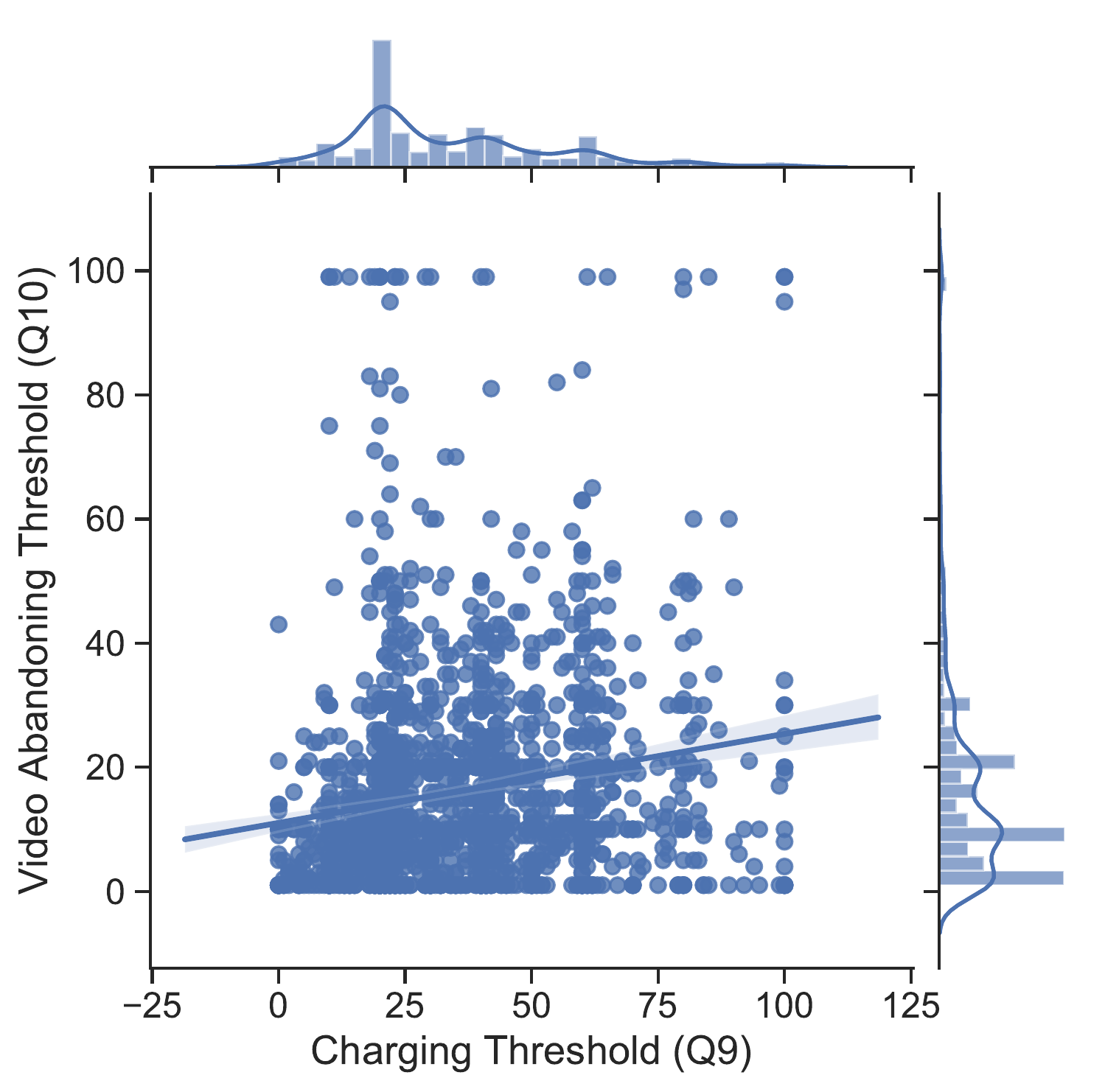}}
 	\subfigure[Zoomed-in density plot.]{
 		\label{fig:videoabandon_corr_kde}
 		\includegraphics[width=0.45\textwidth]{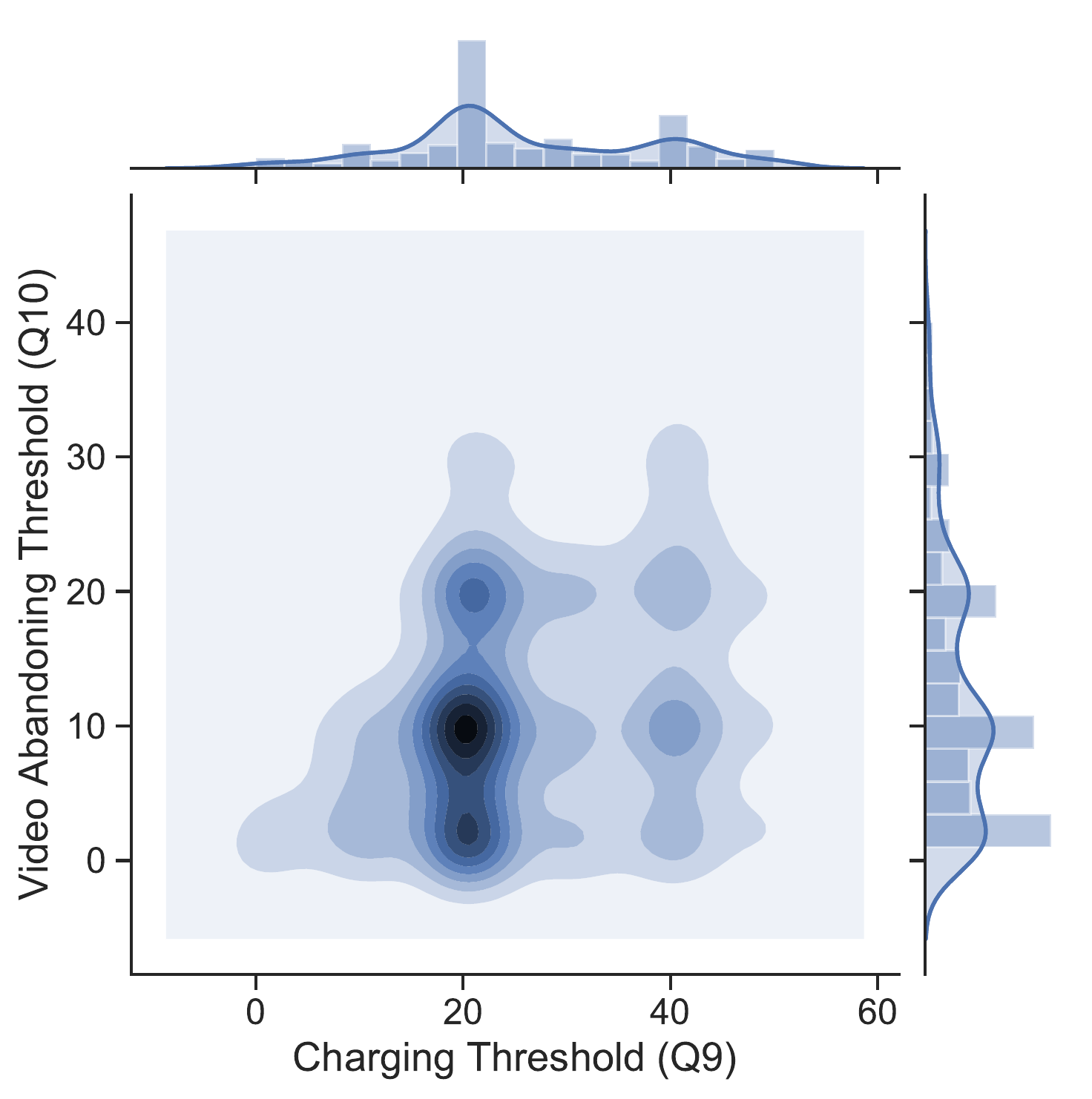}}
 	\caption{(a) The overall regression plot of correlation between charging threshold (Q9) and video abandoning threshold (Q10). (b) The density plot of correlation between charging threshold (Q9) and video abandoning threshold (Q10). A zoomed-in view only for those with charging threshold $\leq 60\%$ and video abandoning threshold $\leq 40\%$.}
 	\label{fig:videoabandon_corr}
\end{figure}

\nop{
\begin{figure}[!h]
\begin{center}
\includegraphics[width=0.45\textwidth]{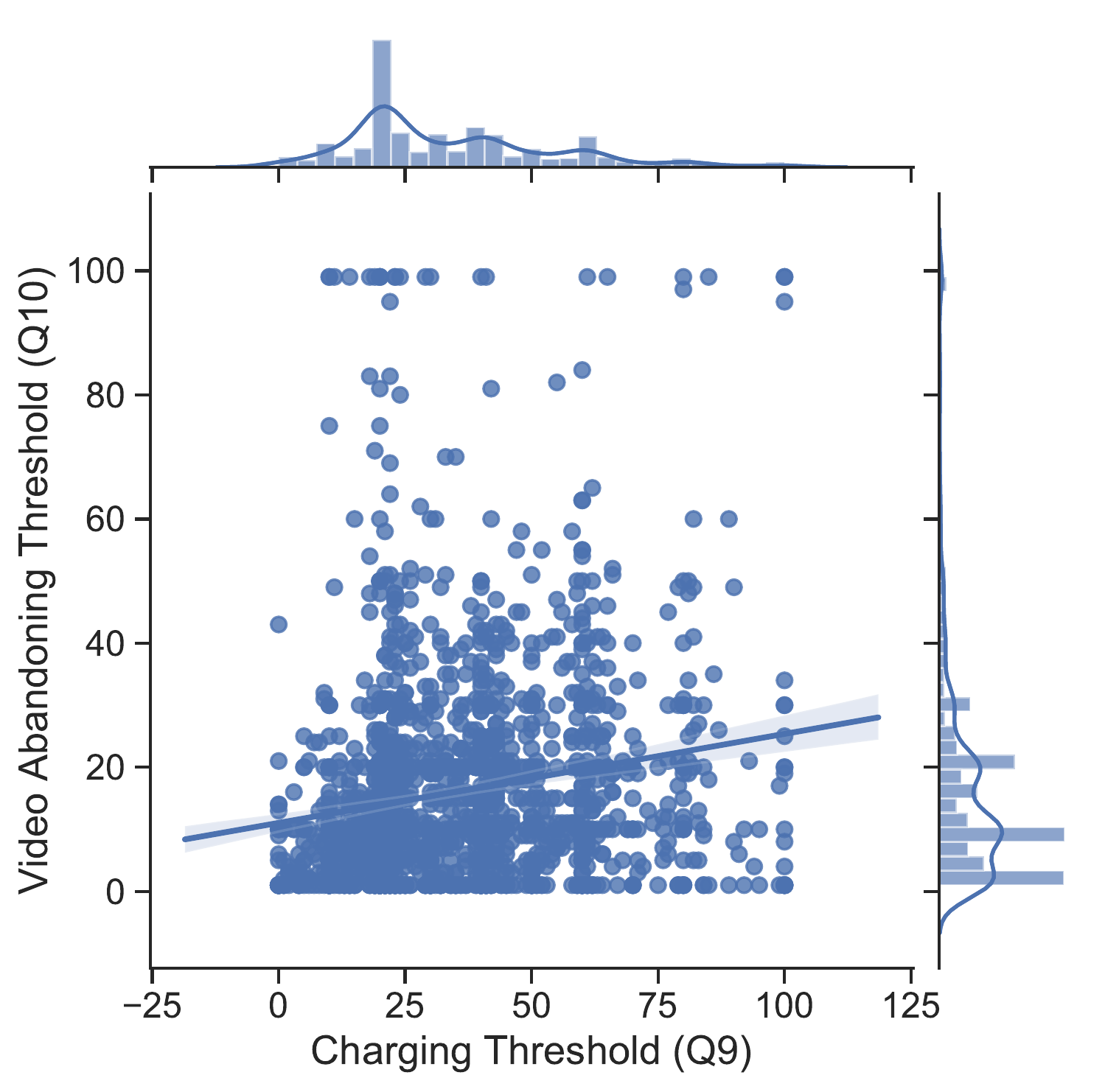}
\caption{The overall regression plot of correlation between charging threshold (Q9) and video abandoning threshold (Q10).}\label{fig:videoabandon_corr}
\end{center}
\end{figure}

\begin{figure}[!h]
\begin{center}
\includegraphics[width=0.45\textwidth]{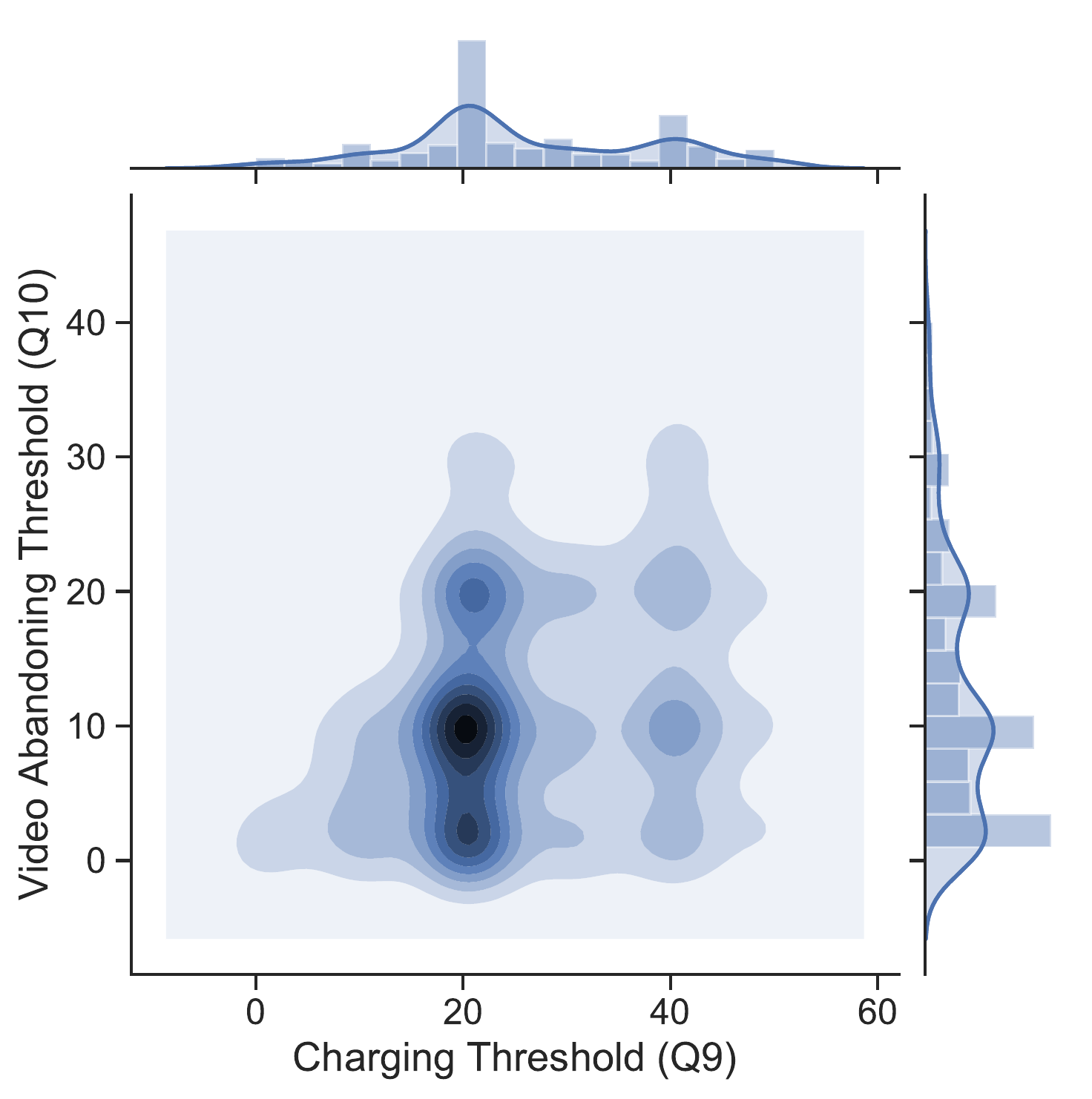}
\caption{The density plot of correlation between charging threshold (Q9) and video abandoning threshold (Q10). A zoomed-in view only for those with charging threshold $\leq 60\%$ and video abandoning threshold $\leq 40\%$.}\label{fig:videoabandon_corr_zoom}
\end{center}
\end{figure}
}

We also perform a correlation analysis between users' feedback of the last two questions: battery charging threshold (Q9) vs. video abandoning threshold (Q10). Fig.~\ref{fig:videoabandon_corr} shows the regression plot and density plot, along with the histograms (and corresponding density contours) at the margins. Based on the figure, we have the following findings:
\begin{itemize}
\item As shown in Fig.~\ref{fig:videoabandon_corr_reg}, across the whole region of battery level from $0$ to $100\%$, there is a weak positive (linear) correlation between users' charging thresholds and their video abandoning thresholds (Pearson's $r=0.18$, \textcolor{MyBlue}{$p < 0.01$})\footnote{Pearson correlation value is widely used to measure the linear correlation between two variables, which has a value region of $[-1, 1]$ with $1$ denoting the strongest positive relationship and $-1$ the negative.}.
\item From the density draw of Fig.~\ref{fig:videoabandon_corr_kde}, we can see that when the battery level is low, a stronger correlation between the two answers can be found (Pearson's $r=0.25$, \textcolor{MyBlue}{$p < 0.01$}). In particular, most users feel the low-battery anxiety (with intention to charge the mobile phone) when the battery level drops to around $20\%$, and thereafter they are likely to give up watching attractive videos for energy saving.
\end{itemize}

\subsection{Advice for Video Streaming Services}

The above observations and findings reveal important information for mobile video streaming services. Besides the traditional QoE metrics of video streaming, e.g., latency and resolution, the energy status of users' mobile phone should also be taken as a critical QoE metric. More attention should be drawn to the mobile users with battery level in the sensible regions, especially those in the extremely sensible region with battery level below $10\%$. If no measures were taken to alleviate the low-battery anxiety, the service provider may lose those customers. Furthermore, if the major potential customers of a video streaming service are mobile users aged above $35$, the provider should take more efforts in alleviating the customers' low-battery anxiety.

There are already some work on energy-aware or low-power video streaming, based on optimization techniques or software/hardware technologies. For example, the display component of the mobile phone consumes a large portion of the total energy, especially when playing videos. Thus, researchers have developed sophisticated schemes to scale the display backlight or transform the video content, so that the display power consumption could be reduced~\cite{cheng2007quality,liu2016gocad}. To further cut down the mobile phone's energy consumption, the computation of video transform work can be mitigated to the cloud or edge servers~\cite{lin2012dynamic,bhojan2018adaptive}. The above energy-aware or low-power video streaming work could alleviate the low-battery anxiety and improve user QoE in mobile video streaming services.

\section{Discussions}\label{sec:discuss}

\subsection{Summary of Major Results}

In this work, to fill the void of thoroughly investigating and quantifying LBA, we conducted a large-scale survey over 2000+ modern mobile users, looked into their reactions towards LBA and quantified their anxiety degree and video abandoning likelihood during the draining of battery power. We have found that, current mobile users are widely affected by the low-battery anxiety, with $91.88\%$ of the surveyed participants claiming to suffer from it more or less. With carefully designed survey questions and tailored answer quantification approaches (e.g., reversed accumulation of charging thresholds), we were able to extract the empirical LBA curve (anxiety degree vs. battery status) and users' video watching behavior curve (video abandoning likelihood vs. battery status). With the quantified LBA and video abandoning likelihood, we provided mobile phone manufactures and mobile OS developers with guidelines in designing low-power or energy-efficient applications and services. In addition, we gave advises to mobile service providers on how the quantified curves and knowledge could be utilized to enhance users' QoE (e.g., in video streaming). By considering both age and gender factors, this quantitative study of LBA sheds insights on the LBA characteristics and reaction likelihood regarding different age and gender groups.

\subsection{How to Alleviate Low-Battery Anxiety?}

There are various ways to alleviate LBA. One is releasing sufficient information about the remaining battery energy or providing accurate and personalized battery lifetime predictions. Another way is deploying widespread charging stations, since when charging is convenient, the mobile users would not feel anxious any more. Nevertheless, this may lead to high expenditure and even wastes on facility investments. Last, most models of mobile phones (with either iOS or android) have the setting to hide the battery (percentage) icon~\cite{remove_battery_icon}. Users who are too sensitive and feel stressed to the draining of batter power might simply use this ``ostrich'' method to alleviate LBA. Efforts can also come from the mobile phone manufacture side, including i) extending the battery life as much as possible, and ii) providing low power working modes to keep the phone connected longer~\cite{iphone_low_power_mode} or new features for more convenient charging like wireless charging.

Potentially, there might be other approaches to dealing with the low-battery anxiety. The essential purpose of this work, however, is to measure and quantify the caused anxiety during the battery power draining. Hence, we much focus on LBA quantification rather than LBA treatment. Our work is expected to provide valuable information, for domain experts like the mobile service providers, the mobile hardware/software designers or even the psychologists, in designing or optimizing solutions to alleviate the LBA accordingly.

\subsection{QoE Quantification and Its Applications}

QoE quantification has been studied for long, but practical measurement and evaluation schemes are still inadequate, due to the difficulties in data collecting, information processing, and knowledge extracting. Besides the design of aforementioned HBI strategies and battery interfaces, quantified QoE model plays an important role in many other fields of the IMWUT community. For example, in~\cite{lee2017user} the authors focused on QoE analysis at the UI level and proposed a QoE analysis tool for mobile applications. Under the context of urban public transport, the authors of~\cite{costa2013railway} presented a case study to investigate the relationship between the users' emotion and experience. 

In other communities, QoE measurement and enhancement are also treated as a critical problem. For example, in the computer networks community, the curve of user's QoE vs. web page response time was leveraged to improve the web service in~\cite{zhang2019e2e}, and the curve was in a similar shape as the LBA curve in this work. A QoE-aware application placement policy was proposed to facilitate service delivery among fog instances and maximize the user QoE in~\cite{mahmud2019quality}. In~\cite{xiao2017qoe}, both user QoE and power efficiency were considered when offloading workload in fog computing networks.

Back to this work, the quantified LBA measure can be a powerful knob in any QoE-aware design and improvement of mobile applications and services. As another supporting evidence, our research group recently took advantage of the quantified LBA for user QoE optimization and designed a low-power video streaming service solution at the Internet edge, in which the quantified LBA plays a critical role in reducing mobile users' anxiety during video streaming.  

\nop{
The technique report of~\cite{techrept_lvss} demonstrates our detailed design (from the video streaming service provider's perspective), as well as how the quantified LBA was utilized to release mobile users' anxiety.}

\subsection{Limitations}
The findings in this work may not apply to situations when charging battery is convenient. We intentionally ignore those situations by explicitly reminding the participants that they were under inconvenient charging situations (Q8). The main reason is that according to previous investigation, battery life is not considered as a ``real problem'' when charging the phone is easy~\cite{hosio2016monetary}. 

Since we distributed our questionnaire over WeChat, the majority of the participants are in China. It is thus unclear whether or not different culture backgrounds may impact the LBA. Nevertheless, we have a good reason to believe that LBA is a world-wide phenomenon, since our findings are consistent with those in the LG survey~\cite{rept_lowbatteryanxiety}. As another limitation, over half ($50.39\%$) of the participants are college students aged between $18$ and $25$. As such, our findings may reflect their LBA situations more accurately, compared to those of other occupations and age groups.

Finally, the results (e.g., extracted anxiety degrees of mobile users) were obtained with survey questions. The underlying assumption for all survey-based studies is that the participants' answers truthfully reflect their feelings and behaviors. This assumption may be challenged, and an alternative method to avoid this pitfall is to directly measure/monitor users' behaviors via wearable sensors~\cite{varkey2012human,visuri2018ubiquitous}. While we value measurement-based studies~\cite{ustev2013user}, it may be difficult to apply this methodology in our context.

\nop{The findings in this work is conditionally reliable, as we constrained the circumstances in our survey questions. For example, in Q10 we explicitly remind the participants that they were under inconvenient charging situations. Thus, the results may not reflect the situation where charging is not a problem. According to previous investigation, when charging the phone is easy, battery life is not considered as a ``real problem''~\cite{hosio2016monetary}.

The conclusions in this work might not be completely applicable to mobile users and markets outside of Chain, as most ($96.56\%$) of the participants are from China. Also, since over half ($50.39\%$) of the participants are college students aged between $18$ and $25$, our findings may reflect their LBA situations more accurately, compared to those of other occupations and age ranges.

The results (e.g., extracted anxiety degrees of mobile users) were obtained in a controlled environment, as all the questions were given under some assumptions and answered in an online form (not a real circumstance). We simply considered the participants' answers regarding their feelings and behaviors as they describe them. That is, their own description of the subjective suffering and reacting was treated as a feasible approach to understanding their feelings and behaviors. This, however, does not necessarily mean that they must have consistent feelings and behaviors in realistic, especially when the contexts are different from what we assumed.}

\section{Ethics}\label{sec:ethics}

Our work involves human subjects, and thus we took care to follow community best practices when conducting our work. Before participating the survey, each mobile user was informed of the intention of this study, the course of data collection and processing, and how the data would be used. The participation is purely voluntary. We only collected data that are necessary for our analysis and quantification of low-battery anxiety, and all the data were anonymized without users' personal identification. Overall, our study would not cause privacy and ethics concerns. \nop{Overall, the field study was performed strictly in a controlled setting and does not raise any ethical issues.}

\section{Conclusions}\label{sec:conclude}

In this paper, we conducted a large-scale survey over $2000+$ mobile users regarding the LBA issue and extracted quantitative models of LBA from the survey data. Through the survey investigation and LBA quantification, we revealed the severity of LBA among modern mobile users, quantified the anxiety degree of mobile users under varying battery levels, and quantified the impacts of LBA on mobile users' video watching behavior. The findings and lessons provided in this work could serve as valuable guidances in enhancing i) the effectiveness of HBI strategies and battery interfaces and ii) the user QoE of mobile applications and services.

\nop{
\begin{acks}
We thank the participants in our study. This work is partly supported by XXX. 
\end{acks}
}

\bibliographystyle{acm}
\bibliography{references}



\appendix

\section{Values of LBA Curve and Video Abandoning Likelihood Curve}
\label{subsec:apdx}

The detailed values of the LBA curve and the video abandoning likelihood curve (along with corresponding battery levels) are provided in Table~\ref{tbl:lba_curve_data} and Table~\ref{tbl:videoabandon_curve_data}, respectively.

\textbf{Note}: To obtain a continuous version of the curve, either parameter regressions (linear/nonlinear) or non-parameter techniques (e.g., B-spline smoothing) can be applied. The piece-wise linear approximation of the curve could be also adopted. As to the curves illustrated in this paper, we did not apply any regression or smoothing techniques, but simply connected points into curves.

\begin{table}[htbp]
  \centering
  \caption{Battery level vs. anxiety degree}\label{tbl:lba_curve_data}
    \begin{tabular}{l rrrrrrrrrr}
    \hline
    \rowcolor{Gray}
    Battery Level (\%) & 1     & 2     & 3     & 4     & 5     & 6     & 7     & 8     & 9     & 10 \\
    Anxiety Degree & 0.9833 & 0.9813 & 0.9788 & 0.9754 & 0.9690 & 0.9656 & 0.9636 & 0.9606 & 0.9557 & 0.9213 \\
    \rowcolor{Gray}
    Battery Level (\%) & 11    & 12    & 13    & 14    & 15    & 16    & 17    & 18    & 19    & 20 \\
    Anxiety Degree & 0.9149 & 0.9070 & 0.9011 & 0.8922 & 0.8804 & 0.8735 & 0.8666 & 0.8538 & 0.8297 & 0.6727 \\
    \rowcolor{Gray}
    Battery Level (\%) & 21    & 22    & 23    & 24    & 25    & 26    & 27    & 28    & 29    & 30 \\
    Anxiety Degree & 0.6314 & 0.5945 & 0.5659 & 0.5438 & 0.5236 & 0.5118 & 0.5069 & 0.5025 & 0.4926 & 0.4523 \\
    \rowcolor{Gray}
    Battery Level (\%) & 31    & 32    & 33    & 34    & 35    & 36    & 37    & 38    & 39    & 40 \\
    Anxiety Degree & 0.4434 & 0.4331 & 0.4247 & 0.4173 & 0.4070 & 0.3996 & 0.3922 & 0.3863 & 0.3735 & 0.3105 \\
    \rowcolor{Gray}
    Battery Level (\%) & 41    & 42    & 43    & 44    & 45    & 46    & 47    & 48    & 49    & 50 \\
    Anxiety Degree & 0.2869 & 0.2702 & 0.2564 & 0.2451 & 0.2352 & 0.2293 & 0.2269 & 0.2224 & 0.2170 & 0.1924 \\
    \rowcolor{Gray}
    Battery Level (\%) & 51    & 52    & 53    & 54    & 55    & 56    & 57    & 58    & 59    & 60 \\
    Anxiety Degree & 0.1850 & 0.1791 & 0.1762 & 0.1727 & 0.1678 & 0.1649 & 0.1634 & 0.1565 & 0.1491 & 0.1097 \\
    \rowcolor{Gray}
    Battery Level (\%) & 61    & 62    & 63    & 64    & 65    & 66    & 67    & 68    & 69    & 70 \\
    Anxiety Degree & 0.0960 & 0.0866 & 0.0807 & 0.0758 & 0.0699 & 0.0659 & 0.0640 & 0.0630 & 0.0610 & 0.0546 \\
    \rowcolor{Gray}
    Battery Level (\%) & 71    & 72    & 73    & 74    & 75    & 76    & 77    & 78    & 79    & 80 \\
    Anxiety Degree & 0.0517 & 0.0512 & 0.0507 & 0.0502 & 0.0492 & 0.0472 & 0.0438 & 0.0438 & 0.0399 & 0.0290 \\
    \rowcolor{Gray}
    Battery Level (\%) & 81    & 82    & 83    & 84    & 85    & 86    & 87    & 88    & 89    & 90 \\
    Anxiety Degree & 0.0261 & 0.0226 & 0.0207 & 0.0177 & 0.0162 & 0.0157 & 0.0153 & 0.0153 & 0.0143 & 0.0133 \\
    \rowcolor{Gray}
    Battery Level (\%) & 91    & 92    & 93    & 94    & 95    & 96    & 97    & 98    & 99    & 100 \\
    Anxiety Degree & 0.0128 & 0.0118 & 0.0113 & 0.0108 & 0.0098 & 0.0098 & 0.0098 & 0.0098 & 0.0089 & 0.0000 \\
    \hline
    \end{tabular}%
\end{table}%

\begin{table}[htbp]
  \centering
  \caption{Battery level vs. video abandoning likelihood (LL.)}\label{tbl:videoabandon_curve_data}
    \begin{tabular}{l rrrrrrrrrr}
    \hline
    \rowcolor{Gray}
    Battery Level (\%) & 1     & 2     & 3     & 4     & 5     & 6     & 7     & 8     & 9     & 10 \\
    Video Abandon LL. & 0.8460 & 0.8282 & 0.8056 & 0.7908 & 0.7362 & 0.7106 & 0.6914 & 0.6654 & 0.6388 & 0.4961 \\
    \rowcolor{Gray}
    Battery Level (\%) & 11    & 12    & 13    & 14    & 15    & 16    & 17    & 18    & 19    & 20 \\
    Video Abandon LL. & 0.4675 & 0.4518 & 0.4395 & 0.4247 & 0.3824 & 0.3716 & 0.3652 & 0.3494 & 0.3307 & 0.2338 \\
    \rowcolor{Gray}
    Battery Level (\%) & 21    & 22    & 23    & 24    & 25    & 26    & 27    & 28    & 29    & 30 \\
    Video Abandon LL. & 0.2131 & 0.2072 & 0.1949 & 0.1855 & 0.1673 & 0.1604 & 0.1565 & 0.1521 & 0.1467 & 0.1147 \\
    \rowcolor{Gray}
    Battery Level (\%) & 31    & 32    & 33    & 34    & 35    & 36    & 37    & 38    & 39    & 40 \\
    Video Abandon LL. & 0.1038 & 0.0999 & 0.0960 & 0.0915 & 0.0881 & 0.0851 & 0.0807 & 0.0778 & 0.0753 & 0.0669 \\
    \rowcolor{Gray}
    Battery Level (\%) & 41    & 42    & 43    & 44    & 45    & 46    & 47    & 48    & 49    & 50 \\
    Video Abandon LL. & 0.0605 & 0.0586 & 0.0541 & 0.0536 & 0.0497 & 0.0477 & 0.0453 & 0.0423 & 0.0394 & 0.0330 \\
    \rowcolor{Gray}
    Battery Level (\%) & 51    & 52    & 53    & 54    & 55    & 56    & 57    & 58    & 59    & 60 \\
    Video Abandon LL. & 0.0300 & 0.0290 & 0.0290 & 0.0281 & 0.0261 & 0.0261 & 0.0261 & 0.0246 & 0.0246 & 0.0212 \\
    \rowcolor{Gray}
    Battery Level (\%) & 61    & 62    & 63    & 64    & 65    & 66    & 67    & 68    & 69    & 70 \\
    Video Abandon LL. & 0.0212 & 0.0207 & 0.0197 & 0.0192 & 0.0187 & 0.0187 & 0.0187 & 0.0187 & 0.0182 & 0.0172 \\
    \rowcolor{Gray}
    Battery Level (\%) & 71    & 72    & 73    & 74    & 75    & 76    & 77    & 78    & 79    & 80 \\
    Video Abandon LL. & 0.0167 & 0.0167 & 0.0167 & 0.0167 & 0.0157 & 0.0157 & 0.0157 & 0.0157 & 0.0157 & 0.0153 \\
    \rowcolor{Gray}
    Battery Level (\%) & 81    & 82    & 83    & 84    & 85    & 86    & 87    & 88    & 89    & 90 \\
    Video Abandon LL. & 0.0143 & 0.0138 & 0.0128 & 0.0123 & 0.0123 & 0.0123 & 0.0123 & 0.0123 & 0.0123 & 0.0123 \\
    \rowcolor{Gray}
    Battery Level (\%) & 91    & 92    & 93    & 94    & 95    & 96    & 97    & 98    & 99    & 100 \\
    Video Abandon LL. & 0.0123 & 0.0123 & 0.0123 & 0.0123 & 0.0113 & 0.0113 & 0.0108 & 0.0108 & 0.0000 & 0.0000 \\
    \hline
    \end{tabular}%
\end{table}%

\nop{
\textcolor{blue}{The piece-wise approximation of the video abandoning likelihood (VAL) curve is drawn in Fig.~\ref{fig:piecewise_val} and corresponding linear functions are shown Table~\ref{tbl:piecewise_val}.}
}

\newpage

\section{Questionnaire of our Survey}\label{sec:apdx_survey}

In our survey, each of the participants was courteously requested to complete ten questions. Note that the original version of the questionnaire was in Chinese, as the voluntary participants are Chinese or comfortable with the language. We translate the questionnaire as follows for the purpose of information sharing.
\\
\\
\fbox{
  \parbox{\textwidth}{%
  \textbf{1. Your gender: }\\
    $\bigcirc$ Male \\
    $\bigcirc$ Female 
  }
}
\\
\fbox{
  \parbox{\textwidth}{%
  \textbf{2. Your age:}\\
$\bigcirc$ Under 18\\
$\bigcirc$ 18 $\sim$ 25\\
$\bigcirc$ 35 $\sim$ 45\\
$\bigcirc$ 45 $\sim$ 65\\
$\bigcirc$ Above 65
  }
}
\\
\fbox{
  \parbox{\textwidth}{%
  \textbf{3. Your occupation:}\\
$\bigcirc$ Student\\
$\bigcirc$ Government/Institute\\		
$\bigcirc$ Company/Corporation\\		
$\bigcirc$ Freelancer\\		
$\bigcirc$ Others
  }
}
\\
\fbox{
  \parbox{\textwidth}{%
  \textbf{4. Your mobile phone brand:}\\
$\bigcirc$ iPhone\\	
$\bigcirc$ Samsung\\		
$\bigcirc$ Huawei\\		
$\bigcirc$ Xiaomi\\		
$\bigcirc$ OPPO\\		
$\bigcirc$ Vivo\\	
$\bigcirc$ Others
  }
}
\\
\fbox{
  \parbox{\textwidth}{%
  \textbf{5. Are you satisfied with the battery capacity of your mobile phone?}\\
$\bigcirc$ Satisfied\\	
$\bigcirc$ Just OK \\		
$\bigcirc$ Not Satisfied
  }
}
\\
\fbox{
  \parbox{\textwidth}{%
  \textbf{6. How many times do you have to charge your mobile phone daily (otherwise it would run out of power)?}\\
$\bigcirc$ $\leq$ 1 \\		
$\bigcirc$ 2 \\		
$\bigcirc$ 3 \\		
$\bigcirc$ 4 \\		
$\bigcirc$ $>$ 4 
  }
}
\\
\fbox{
  \parbox{\textwidth}{%
  \textbf{7. The frequency of your daily use of portable power bank (or backup battery):}\\
$\bigcirc$ Never use \\
$\bigcirc$ Occasionally use \\
$\bigcirc$ Frequently use
  }
}
\\
\fbox{
  \parbox{\textwidth}{%
  \textbf{8. When inconvenient to charge the mobile phone, will you suffer from anxiety or panic when the battery level is low (say around $20\%$)?}\\
  $\bigcirc$ Not at all \\
  $\bigcirc$ A little \\
  $\bigcirc$ Confirmed suffering \\
  $\bigcirc$ Severely suffering 
  }
}
\\
\fbox{
  \parbox{\textwidth}{%
  \textbf{9. At what battery level (in percentage from 0 to 100\%) will you charge the mobile phone, when it is possible?}\\
\\
\includegraphics[scale=0.5]{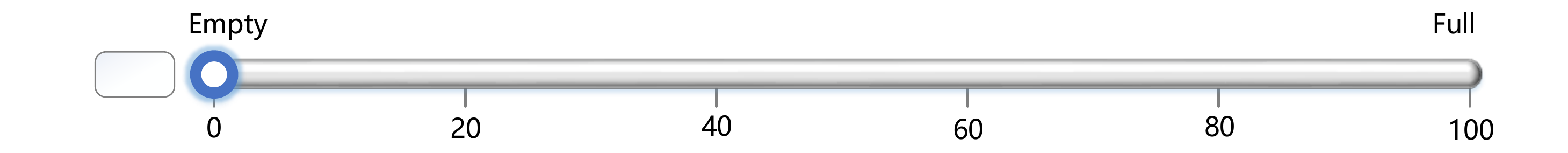}

\begin{flushright}
\textit{*You can drag the slider to select a value or enter the value directly in the left box.}
\end{flushright}
  }
}
\\
\fbox{
  \parbox{\textwidth}{%
  \textbf{10. At what battery level (in percentage from 1\% to 100\%) will you give up watching a video you are interested in, when you are browsing the \emph{WeChat Moment} or \emph{Weibo}?}\\
\\
\includegraphics[scale=0.5]{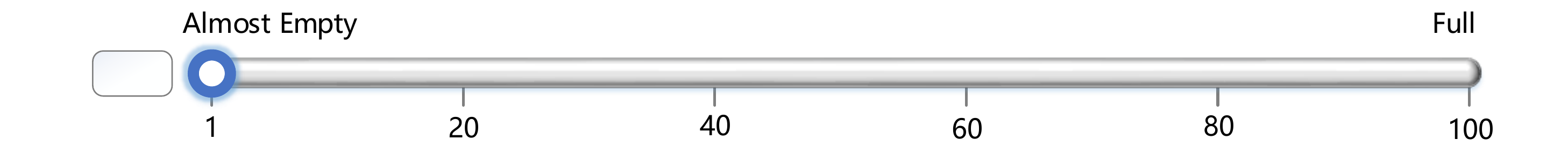}

\begin{flushright}
\textit{*You can drag the slider to select a value or enter the value directly in the left box.}
\end{flushright}
  }
}
\\

\end{document}